\documentclass[useAMS,usenatbib]{mn2e}
\usepackage{graphicx}
\usepackage{aas_macros}

\newcommand{\ltappeq}{\raisebox{-0.6ex}{$\,\stackrel
{\raisebox{-.2ex}{$\textstyle <$}}{\sim}\,$}}
\newcommand{\gtappeq}{\raisebox{-0.6ex}{$\,\stackrel
{\raisebox{-.2ex}{$\textstyle >$}}{\sim}\,$}}

\title[Space density of magnetic CVs]{The space density of magnetic cataclysmic variables}

\author[M.L. Pretorius, C. Knigge and A.D. Schwope]{Magaretha L. Pretorius,$^{1,2}$\thanks{E-mail: m.pretorius@soton.ac.uk (MLP); c.knigge@soton.ac.uk (CK); aschwope@aip.de (ADS)} Christian Knigge$^{2}$\footnotemark[1] and Axel D. Schwope$^{3}$\footnotemark[1]\\
$^{1}$European Southern Observatory, Alonso de C\'{o}rdova 3107, Vitacura, Santiago, Chile\\
$^{2}$School of Physics and Astronomy, University of Southampton, Highfield, Southampton SO17 1BJ, United Kingdom\\
$^{3}$Leibniz-Institut f{\"u}r Astrophysik Potsdam, An der Sternwarte 16, 14482 Potsdam, Germany\\}

\begin{document}


\pagerange{\pageref{firstpage}--\pageref{lastpage}} \pubyear{}

\maketitle

\label{firstpage}

\begin{abstract}
We use the complete, X-ray flux-limited \emph{ROSAT} Bright Survey
(RBS) to measure the space density ($\rho$) of magnetic cataclysmic
variables (mCVs). The survey provides complete optical identification
of all sources with count rate $> 0.2\,\mathrm{s^{-1}}$ (corresponding
to $F_X \ga 2 \times 10^{-12}\,\mathrm{erg\,cm^{-2}s^{-1}}$) over half
the sky ($|b|>30^\circ$), and detected 6 intermediate polars (IPs) and
24 polars.  If we assume that the 30 mCVs included in the RBS are
representative of the intrinsic population, the space density of mCVs
is $8^{+4}_{-2} \times 10^{-7}\,\mathrm{pc^{-3}}$. Considering polars
and IPs separately, we find $\rho_{polar}=5^{+3}_{-2} \times
10^{-7}\,\mathrm{pc^{-3}}$ and $\rho_{IP}=3^{+2}_{-1} \times
10^{-7}\,\mathrm{pc^{-3}}$.  Allowing for a 50\% high-state duty cycle
for polars (and assuming that these systems are below the RBS
detection limit during their low states) doubles our estimate of
$\rho_{polar}$ and brings the total space density of mCVs to
$1.3^{+0.6}_{-0.4} \times 10^{-6}\,\mathrm{pc^{-3}}$.  We also place
upper limits on the sizes of faint (but persistent) mCV populations
that might have escaped detection in the RBS.  Although the large
uncertainties in the $\rho$ estimates prevent us from drawing strong
conclusions, we discuss the implications of our results for the
evolutionary relationship between IPs and polars, the fraction of CVs
with strongly magnetic white dwarfs (WDs), and for the contribution of
mCVs to Galactic populations of hard X-ray sources at $L_X \ga 10^{31}
{\rm erg s^{-1}}$. Our space density estimates are consistent with the
very simple model where long-period IPs evolve into polars and account
for the whole short-period polar population. We find that the fraction
of WDs that are strongly magnetic is not significantly higher for CV
primaries than for isolated WDs. Finally, the space density of IPs is
sufficiently high to explain the bright, hard X-ray source population
in the Galactic Centre.
\end{abstract}

\begin{keywords}
binaries -- stars: cataclysmic variables, -- X-rays: binaries -- methods: observational, statistical.
\end{keywords}

\section{Introduction}
\label{sec:intro}
Cataclysmic variables (CVs) are interacting binary stars in which a
white dwarf (WD) accretes matter from a low-mass, Roche-lobe-filling
companion. In around 20\% of known CVs, the magnetic field of the WD
is sufficiently strong to control at least the inner part of the
accretion flow; these magnetic CVs (mCVs) are divided into two
classes, namely polars and intermediate polars (IPs). The defining
property of a polar is that the WD is locked in synchronous rotation
with the binary orbit, while IPs have WD spin periods typically much
shorter than the orbital period ($P_{orb}$).

\begin{figure}
 \includegraphics[width=84mm]{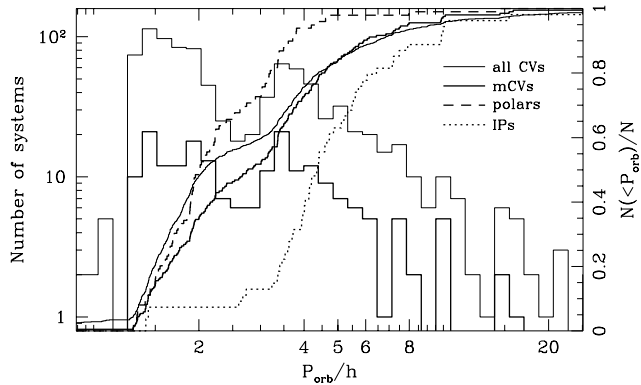} 
  \caption {The orbital period distribution of all CVs (fine histogram), and mCVs (bold histogram). Cumulative distributions are also shown for polars (dashed histogram) and IPs (dotted histogram). Almost all IPs are found at long $P_{orb}$, while most polars are short-$P_{orb}$ systems. Although the period distribution of mCVs shows a period gap, it is less pronounced than for non-magnetic CVs. The periods used here are taken from version 7.18 of the catalogue of Ritter \& Kolb (2003).
}
 \label{fig:perhis}
\end{figure}

In many ways, the formation and evolution of magnetic and non-magnetic
CVs is thought to be similar. Both types of systems form through
common envelope (CE) evolution, evolve first from long to short
$P_{orb}$ because of angular momentum loss (AML), and eventually
experience period bounce, when the thermal time-scale of the donor
becomes longer than its mass loss time-scale. In fact, the main
proposed difference between the evolution of mCVs and non-magnetic CVs
affects only the polars. In these systems, magnetic braking (MB),
which is thought to be the dominant AML mechanism for most CVs above
the period gap, is likely to be suppressed
(e.g.\ \citealt{LiWickramasinghe98, TownsleyGansicke09}). The
$P_{orb}$ distribution of mCVs is broadly in line with these ideas: if
polars and IPs are considered jointly, their $P_{orb}$ distribution is
very similar to that of non-magnetic CVs, showing both a period gap in
the range $2\,{\rm hr} \ltappeq P_{orb} \ltappeq 3\,{\rm hr}$ and a
period minimum at around $P_{orb} \simeq 80\,{\rm min}$ (see
Fig.~\ref{fig:perhis}).

However, there are at least three important open questions concerning the 
formation, evolution and Galactic abundance of mCVs: 
\begin{itemize} 
\item[(i)] Is there an evolutionary relationship between IPs and
polars?
\item[(ii)] What is the intrinsic fraction of mCVs amongst the general
CV population, and can this be reconciled with the incidence of
magnetic WDs in the isolated WD population? 
\item[(iii)] Do mCVs dominate the total Galactic X-ray source
populations above $L_X \sim 10^{31}\,{\rm erg s^{-1}}$?
\end{itemize}
In principle, all of these questions can be addressed empirically, but
this requires reliable measurements of the space density of IPs and
polars. The goal of the present work is to provide such measurements.

The plan of this paper is as follows. In Section~\ref{sec:context}, we
will provide some additional background on the questions listed above,
and discuss previous attempts to measure the space densities of polars
and IPs. Our flux-limited, X-ray-selected mCV sample is presented in
Section~\ref{sec:sample}, with distance and X-ray luminosity estimates
discussed in Section~\ref{sec:dist} and \ref{sec:xlum}.  In
Section~\ref{sec:rhocalc}, we describe how we use this sample to
calculate space densities, and present our results. The results are
discussed in Section~\ref{sec:discussion}, and, finally, we list our
conclusions in Section~\ref{sec:concl}.

\section{Context}
\label{sec:context}

\subsection{The relationship between intermediate polars and polars}

It has been known for a long time that most IPs are found above the
period gap and most polars below (see Fig.~\ref{fig:perhis}). This
immediately suggests that IPs may evolve into polars
(\citealt{ChanmugamRay84,KingFrankRitter85}).  This is a physically
appealing idea.  Whether the WD in a given system will synchronize
depends on the strength of the WD's magnetic field, the orbital
separation, and the mass-transfer rate ($\dot{M}$) from the secondary
star (with large magnetic field, small orbital separation, and low
$\dot{M}$ favouring synchronization).  Since MB drives much higher
mass-transfer rates above the period gap than gravitational radiation
(GR) does below, it is plausible that many accreting magnetic WDs may
only achieve synchronization once they have crossed the period gap.

There is, however, a serious problem with this attractive scenario. If
most IPs evolve into polars, and most polars are the descendants of
IPs, their field strengths should be comparable. Yet, empirically, the
magnetic fields of the WDs in IPs ($B_{IP} \la 10 {\rm MG}$) are
systematically weaker than those of the WDs in polars ($B_{polar} \sim
10 - 100 {\rm MG}$)
(e.g.\ \citealt{ChanmugamRay84,WickramasingheWuFerrario91,LambMelia87,SchmidtSzkodySmith96,SchmidtFerrarioWickramasinghe01,NortonWynnSomerscales04,SchwopeSchreiberSzkody06,Butters09a}).

There are several possible resolutions to this problem. Perhaps the
simplest (in terms of binary evolution) is that the high accretion
rates in IPs may partially ``bury'' the WD magnetic fields, so that
the observationally inferred field strengths for these systems are
systematically biased low \citep{Cumming02}. Alternatively,
\cite{Patterson94} points out that it may not be necessary for all (or
even most) long-period IPs to evolve into polars in order to account
for the entire short-period polar population, since the evolutionary
time-scales are so much shorter above the period gap than below. This
idea still requires an explanation for the fate of the remaining
long-period IPs, although perhaps they simply become unobservable as
their accretion flows become clumpy and their luminosity is pushed
into the EUV band \citep{WickramasingheWuFerrario91}. Finally, the
opposite view would be that the short-period polar population may be
dominated by systems born below the period gap. But in this case,
there would have to be a direct link between the WD field strength and
the orbital period after emergence from the CE phase
(e.g.\ \citealt{Tout08}).

One way to shed light on the relationship between IPs and polars is
via their respective space densities. For example, if all long-period
IPs evolve into short-period polars, and all short-period polars are
the progeny of long-period IPs, then their space densities should be
proportional to the evolutionary time-scale associated with these
phases. In this particular example, we would predict that
$\rho_{polar}/\rho_{IP} \simeq \tau_{GR}/\tau_{MB} >> 1$.

\subsection{The intrinsic fraction of magnetic CVs}

Magnetic systems make up $\simeq 20\%$ of the known CV population
\citep{rkcat}.  At first sight, this is a surprisingly high fraction,
given that the strong magnetic fields characteristic of IPs and polars
($B \gtappeq 10^6\,{\rm G}$) are found in only $\simeq 10\%$ of
isolated WDs (e.g.\ \citealt{Kawka07,Kulebi09}).  If these numbers are
representative of the intrinsic incidence of magnetism amongst CVs and
single WDs, the difference between them would have significant
implications: either strong magnetic fields would have to favour the
production of CVs, or some aspect(s) of pre-CV evolution would have to
favour the production of strong magnetic fields.

A model along the latter lines was recently proposed by
\cite{Tout08}. They argue that all strongly magnetic WDs are the
product of binary evolution, and that the observed strong fields
develop during the CE phase. In this picture, mCVs are associated with
CE events that produce very close binary systems (which are then
brought into contact by MB or GR), while isolated magnetic WDs are
produced when two degenerate cores undergo a complete merger during
the CE phase.

However, it is actually by no means clear yet that magnetism is really
more common in CV primaries than in isolated WDs. The main problem is
that the observed fraction of magnetic systems amongst known CVs is
almost certainly affected by serious selection biases. For example,
since mCVs are known to be relatively X-ray bright, they are likely to
be over-represented in X-ray-selected samples. Conversely, polars, in
particular, are relatively faint in optical light (since they do not
contain optically bright accretion disks), so they are likely to be
under-represented in optically-selected samples. Given that the
overall CV sample is a highly heterogeneous mixture of X-ray-,
optical- and variability-selected sub-samples (which also usually lack
clear flux limits), it is very difficult to know how the observed
fraction of mCVs relates to the intrinsic fraction of magnetic WDs in
CVs.

\subsection{Galactic X-ray Source Populations}

There have been many attempts to determine the make-up and luminosity
function of Galactic X-ray source populations in a variety of
environments, ranging from the Milky Way as a whole
(e.g. \citealt{Sazonov06}), the Galactic Centre
(e.g. \citealt{Muno04,Muno06,Muno09}), the Galactic Ridge
(e.g.\ \citealt{Revnivtsev06,Krivonos07,Hong12}), and even globular
clusters (e.g. \citealt{Heinke05}). Remarkably, in all of these
environments, mCVs have been proposed as the dominant population of
X-ray sources above $L_X \gtappeq 10^{31}{\rm erg s^{-1}}$.

However, in most of the above studies, the breakdown of the observed
X-ray source samples into distinct populations is subject to
considerable uncertainty. In general, only a small fraction of the
sources have optical counterparts and/or properties that would permit
a clear classification. Thus, for the most part, identifications of
observed sources with physical populations have to rely on gross
properties (such as X-ray colours and luminosities) and statistical
comparisons of observed and expected number counts. The local space 
densities of the relevant populations are arguably the most important
ingredient in these comparisons. In effect, the question being asked
is whether the extrapolation of the local space density to the
environment being investigated can account for the observed number of
sources seen there. In the case of mCVs, such extrapolations are
difficult, primarily because the local space densities are rather
poorly constrained. 

\subsection{The Space Density of Magnetic CVs: Previous Work}

\subsubsection{Polars} 

Estimates for the space density of polars include $\sim 3.5 \times
10^{-7}\,\mathrm{pc^{-3}}$ \citep{Patterson84}, $\sim 3 \times
10^{-7}\,\mathrm{pc^{-3}}$ \citep{bible}, $\sim 5 \times
10^{-7}\,\mathrm{pc^{-3}}$ \citep{Cropper90}, and $1.3 \times
10^{-6}\,\mathrm{pc^{-3}}$ \citep{Araujo-BetancorGansickeLong05}. As
emphasized particularly in the last of these studies, all of these
numbers are, strictly speaking, lower limits, because they are
effectively based on the number of systems observed within a
particular distance cut-off.

\cite{ThomasBeuermann98} derive an estimate of $\rho_{polar} \simeq
6.1 \times 10^{-7}\,\mathrm{pc^{-3}}$ from the sample of polars
detected by \emph{ROSAT} (see also
\citealt{BeuermannSchwope94}). However, they also note that, allowing
for incompleteness, the space density could be as high as $1.2 \times
10^{-6}\,\mathrm{pc^{-3}}$ for short-period polars and $1.7 \times
10^{-7}\,\mathrm{pc^{-3}}$ for long-period polars. \cite{Ramsay04}
discuss the issue of incompleteness in observed samples, particularly
that arising from the extended low-states exhibited by polars.

\subsubsection{Intermediate Polars}

Fewer estimates are available for the space density of
IPs. \cite{bible} gives $\rho_{IP} \sim 3 \times
10^{-8}\,\mathrm{pc^{-3}}$, while \cite{Revnivtsev08} find $\rho_{IP}
\simeq 1.2 \times 10^{-7}\,\mathrm{pc^{-3}}$, from a sample of CVs
detected by \emph{INTEGRAL}.  The scarcity of space density estimates
for IPs is particularly problematic, because it is IPs, rather than
polars, which are often assumed to be the dominant bright hard X-ray
source population in many Galactic environments (see
Section~\ref{sec:xraypop}).

\section{The flux-limited CV sample}
\label{sec:sample}

Given that we do not yet have a useful volume-limited CV sample, the
most suitable sample to use for statistical studies is a purely
flux-limited one (e.g.\ \citealt{PretoriusKniggeKolb07}). 

In two previous papers (\citealt{NEPrho,nonmagphi}) we presented
constraints on the space density and X-ray luminosity function of
non-magnetic CVs, based on the \emph{ROSAT} Bright Survey (RBS;
\citealt{Schwope00}), as well as on the deeper, but smaller area
\emph{ROSAT} North Ecliptic Pole (NEP) survey
(e.g.\ \citealt{Gioia03}; \citealt{Henry06}).  Here, we use the RBS to
derive the corresponding observational constraints on the mCV
population.

While several non-magnetic CVs are known to have very faint X-ray
luminosities, mCVs are expected to be intrinsically brighter in
X-rays, because of the almost radial accretion flow (polars in
particular also have higher soft X-ray to optical flux ratios than
non-magnetic CVs, explaining why so many where first discovered by
\emph{ROSAT}; see fig.~2 of \citealt{BeuermannThomas93}). This means
that the bright flux limit of the RBS is not as important a concern
when studying mCVs.

The RBS is a flux-limited part of the \emph{ROSAT} All-Sky Survey
(RASS; see \citealt{rosatbsc} and \citealt{rosatfsc}), consisting of
bright (count rate $>0.2\,\mathrm{s^{-1}}$), high Galactic latitude
($|b|>30^\circ$) sources. It has complete optical follow-up
(\citealt{Schwope00}; \citealt{SchwopeBrunnerBuckley02}), and includes
30 mCVs --- 6 IPs and 24 polars. Assuming a $30\,\mathrm{keV}$ thermal
bremsstrahlung spectrum, and a $30\,\mathrm{eV}$ blackbody spectrum
(see Section~\ref{sec:xlum}), both with
$N_H=10^{20}\,\mathrm{cm^{-2}}$, the limiting count rate corresponds
to flux limits of $F_X \ga 3 \times
10^{-12}\mathrm{erg\,cm^{-2}s^{-1}}$ and $2 \times
10^{-12}\mathrm{erg\,cm^{-2}s^{-1}}$, respectively, in the
0.12--2.48~keV band .

In Table~\ref{tab:distances} we list the 30 mCVs that make up our
flux-limited sample. It has been suggested that two more CVs detected
in the RBS are magnetic; these are TW Pic (see
\citealt{Mouchet91,PattersonMoulden93,Norton00}) and V405 Peg
\citep{Thorstensen09}. We do not include these two systems here, since
there is no conclusive evidence of a magnetic nature for either of
them. On the other hand, CC Scl was until recently classified as a
normal SU UMa star, but has now been shown to be an IP
\citep{Woudt12}; we therefore include it in our mCV
sample\footnote{Although we previously incorrectly included CC Scl in
  the RBS non-magnetic CV sample, it made only a very small
  contribution to our earlier estimate of the non-magnetic CV space
  density \citep{nonmagphi}, so that the result presented there was
  not significantly affected. It is however important to include CC
  Scl here, because the total space density of magnetic systems is
  smaller.}.

\begin{table*}
 \centering
 \begin{minipage}{168mm}
  \caption{The 30 mCVs detected in the RBS, together with their subtype (polar or IP), orbital periods, Galactic latitudes, \emph{ROSAT} PSPC count rates, distances, and X-ray luminosities. The fraction of the total mCV space density contributed by each system ($\rho_j/\rho_0$) is given in the 8th column (this ignores all errors, although uncertainties are correctly accounted for later; see Section~\ref{sec:calc}). References are for the classification as a magnetic CV, $P_{orb}$, and published distances, where available. Note that the distance and $L_X$ estimates for CD Ind, IW Eri, CV Hyi, and FH UMa are very uncertain (see Section~\ref{sec:problems} and \ref{sec:sensdistance}).}
  \label{tab:distances}
  \begin{tabular}{@{}llllllllll@{}}
  \hline
System  &RBS\# & Type & $P_{orb}/\mathrm{h}$ & $b$ & count rate$/\mathrm{s^{-1}}$& $d/\mathrm{pc}$ & $\mathrm{log}(L_X/\mathrm{erg\,s^{-1}})$ & $\rho_j/\rho_0$ & References \\
 \hline
CC Scl   &1969&IP& 1.402& $-68.7^\circ$ &0.28(6) & $200^{+110}_{-70}$  & $31.3(4)$             & 0.089& 1,2,3         \\[0.1cm]
EX Hya   &1173&IP& 1.638& $+33.6^\circ$ & 5.4(3)  & $64.5 \pm 1.2$   & $31.59(4)$            & 0.043& 4,5           \\[0.1cm]
AO Psc   &1914&IP& 3.591& $-53.3^\circ$ & 0.37(4) & $330^{+180}_{-120}$ & $32.2(4)$             & 0.048& 6,7           \\[0.1cm]
DO Dra   &1022&IP& 3.969& $+44.5^\circ$ & 0.59(3) & $155 \pm 35$     & $31.4(2)$             & 0.119& 8,9           \\[0.1cm]
TV Col   & 655&IP& 5.486& $-30.6^\circ$ & 0.36(4) & $370^{+17}_{-15}$   &$32.22^{+0.08}_{-0.09}$   & 0.046& 10,11         \\[0.1cm]
EI UMa   & 713&IP& 6.434& $+37.4^\circ$ & 0.57(4) & $750^{+100}_{-200}$ & $33.2^{+0.1}_{-0.3}$     & 0.021& 12,13         \\[0.1cm]
CV Hyi   & 213&AM& 1.297& $-50.7^\circ$ & 0.28(4) & $550^{+450}_{-250}$ & $32.3(6)$             & 0.013& 14            \\[0.1cm]
V4738 Sgr&1678&AM& 1.300& $-33.9^\circ$ & 0.33(4) & $250^{+90}_{-70}$   & $31.8^{+0.4}_{-0.5}$     & 0.035& 14            \\[0.1cm]
EV UMa   &1219&AM& 1.328& $+63.1^\circ$ & 1.86(6) & $690^{+270}_{-190}$ & $33.3(3)$             & 0.003& 15,16         \\[0.1cm]
GG Leo   & 842&AM& 1.331& $+49.0^\circ$ & 1.1(1)  & $170^{+90}_{-60}$   & $31.5(4)$             & 0.058& 17,18         \\[0.1cm]
FH UMa   & 904&AM&1.336:& $+48.4^\circ$ & 0.26(2) & $590^{+480}_{-270}$ & $32.0(6)$             & 0.023& 19            \\[0.1cm]
EF Eri   & 398&AM& 1.350& $-57.4^\circ$ & 6.2(3)  & $163^{+66}_{-50}$   & $32.0^{+0.3}_{-0.4}$     & 0.023& 20,21         \\[0.1cm]
IW Eri   & 541&AM&1.452:& $-40.6^\circ$ & 0.23(3) & $270^{+230}_{-120}$ & $31.6(6)$             & 0.048& 22            \\[0.1cm]
EU UMa   &1039&AM& 1.502& $+76.3^\circ$ & 3.3(2)  & $240^{+90}_{-60}$   & $32.4(3)$             & 0.010& 18,23,24      \\[0.1cm] 
EQ Cet   & 206&AM& 1.547& $-80.9^\circ$ & 0.34(3) & $270^{+130}_{-90}$  & $31.4(4)$             & 0.077& 22,25         \\[0.1cm]
V393 Pav &1664&AM& 1.647& $-31.3^\circ$ & 0.90(7) & $340^{+170}_{-110}$ & $32.4(5)$             & 0.011& 26            \\[0.1cm]
EG Lyn   & 696&AM& 1.656& $+34.5^\circ$ & 0.25(3) & $470^{+230}_{-160}$ & $32.4(4)$             & 0.010& 22,27         \\[0.1cm]
RS Cae   & 599&AM&1.699:& $-39.1^\circ$ & 1.20(8) & $880^{+330}_{-240}$ & $33.0(3)$             & 0.004& 28            \\[0.1cm]      
CD Ind   &1735&AM& 1.848& $-41.4^\circ$ & 0.38(4) & $350^{+140}_{-100}$ & $31.7(4)$             & 0.044& 29,30         \\[0.1cm]
BL Hyi   & 232&AM& 1.894& $-48.6^\circ$ & 2.8(2)  & $163^{+18}_{-26}$   & $31.7(2)$             & 0.037& 31,32,33,34,35\\[0.1cm]
EK UMa   & 911&AM& 1.909& $+55.2^\circ$ & 1.11(5) & $590^{+340}_{-210}$ & $32.6(4)$             & 0.008& 36,37,38      \\[0.1cm]
AN UMa   & 938&AM& 1.914& $+62.1^\circ$ & 1.77(7) & $300^{+150}_{-100}$ & $31.9(4)$             & 0.029& 39,40,41      \\[0.1cm]
V1007 Her&1646&AM& 1.999& $+33.3^\circ$ & 0.24(2) & $560^{+280}_{-180}$ & $32.3(5)$             & 0.011& 42            \\[0.1cm]
HU Aqr   &1724&AM& 2.084& $-32.6^\circ$ & 0.81(7) & $290^{+110}_{-80}$  & $32.2^{+0.4}_{-0.5}$     & 0.015& 43,44,45      \\[0.1cm]
UW Pic   & 658&AM& 2.223& $-32.7^\circ$ & 0.73(9) & $300 \pm 60$     & $31.9(3)$             & 0.028& 46,47         \\[0.1cm]
RX J0859 & 734&AM&2.397& $+30.9^\circ$ & 0.23(3) & $530^{+260}_{-180}$ & $32.6(4)$             & 0.008& 48,49,50      \\[0.1cm]
CW Hyi   & 324&AM& 3.030& $-45.9^\circ$ & 0.26(4) & $500^{+250}_{-170}$ & $32.4(4)$             & 0.035& 22            \\[0.1cm]
RX J1610 &1563&AM&3.176& $+37.3^\circ$ & 0.36(4) & $380^{+190}_{-120}$ & $32.1(5)$             & 0.050& 22,51         \\[0.1cm]
1RXS J231603&1973&AM&3.491& $-58.6^\circ$ & 1.10(9) & $460^{+170}_{-120}$ & $32.8(3)$             & 0.026& 51,52         \\[0.1cm]
AI Tri   & 274&AM& 4.602& $-30.3^\circ$ & 0.54(5) & $620 \pm 100$    & $32.5(3)$             & 0.031& 53            \\[0.1cm]
 \hline
 \end{tabular}
\\
References: 
1. \cite{Woudt12};
2. \cite{ChenODonoghueStobie01};
3. \cite{TappertAugusteijnMaza04};
4. \cite{Sterken93};
5. \cite{Beuermann03};
6. \cite{PattersonPrice81};
7. \cite{Patterson84};
8. \cite{MateoSzkodyGarnavich91};
9. \cite{HoardLinnellSzkody05};
10. \cite{Hellier93};
11. \cite{McArthur01};
12. \cite{Reimer08};
13. \cite{Thorstensen86};
14. \cite{BurwitzReinschBeuermann97};
15. \cite{OsborneBeardmoreWheatley94};
16. \cite{Katajainen00};
17. \cite{Burwitz98};
18. \cite{RamsayCropperMason04};
19. \cite{Singh95};
20. \cite{Thorstensen03};
21. \cite{Williams79};
22. \cite{SchwopeBrunnerBuckley02};
23. \cite{Mittaz92};
24. \cite{Howell95};
25. \cite{SchwopeSchwarzGreiner99};
26. \cite{ThomasBeuermannSchwope96};
27. \cite{CaoWeiHu99};
28. \cite{BurwitzReinschSchwope96};
29. \cite{SchwopeBuckleyODonoghue97};
30. \cite{Ramsay99};
31. \cite{Araujo-BetancorGansickeLong05};
32. \cite{BeuermannSchwopeWeissieker85};
33. \cite{Visvanathan84};
34. \cite{Beuermann00};
35. \cite{Glenn94};
36. \cite{Morris87};
37. \cite{ClaytonOsborne94};
38. \cite{BeuermannDiesePaik09};
39. \cite{KrzeminskiSerkowski77};
40. \cite{Bonnet-Bidaud96};
41. \cite{Liebert82};
42. \cite{GreinerSchwarzWenzel98};
43. \cite{SchwopeThomasBeuermann93};
44. \cite{Ciardi98};
45. \cite{SproatsHowellMason96}.
46. \cite{ReinschBurwitzBeuermann94};
47. \cite{Romero-Colmenero03};
48. \cite{BeuermannBurwitz95};
49. \cite{BeuermannThomasReinsch99};
50. \cite{Gansicke09};
51. \cite{Rodrigues06};
52. \cite{BeuermannThomas93};
53. \cite{Schwarz98}.\hfill
\end{minipage}
\end{table*}

\section{Distance estimates}
\label{sec:dist}
We require distance estimates for all mCVs in the sample. Where
possible, we use published distances, based on either trigonometric
parallax, or photometric parallax of the WD or donor star. In most
cases, however, the best distance estimates we can obtain are quite
uncertain. This will be the most important limit on the precision of
our space density measurement. The estimates we use are listed in
Table~\ref{tab:distances}, and we provide more information below.

\subsection{Systems with reliable distance estimates}

\subsubsection{Trigonometric parallax}
The IPs EX Hya and TV Col, and the polar EF Eri have very reliable
distance estimates from parallax measurements (\citealt{McArthur01};
\citealt{Beuermann03}; \citealt{Thorstensen03}).

\subsubsection{Photometric parallax}
\label{sec:photpardist}
Several more systems have reliable distance estimates from photometric
parallax of one of the stellar components. These are based on, e.g.,
the detection of donor star features in the optical or near-IR
spectrum (from which the donor's contribution to the total flux in a
given band can be estimated), or FUV observations in a low state,
where the WD dominates the flux.

Based on donor star features in the near-IR spectrum of DO Dra,
\cite{MateoSzkodyGarnavich91} find a distance of
$155\pm35\,\mathrm{pc}$. This is very close agreement with an estimate
from the FUV detection of the WD \citep{HoardLinnellSzkody05}.

\cite{Araujo-BetancorGansickeLong05} find a distance of
$163^{+18}_{-26}$ for BL Hyi, from a FUV detection of the WD (several
more estimates, all consistent with this, have been derived from
detections of the donor star; see \citealt{Visvanathan84},
\citealt{BeuermannSchwopeWeissieker85}, \citealt{Glenn94},
\citealt{Beuermann00}).

HU Aqr is a deeply eclipsing polar; \cite{Ciardi98} find $K=15.3(1)$
in eclipse, and note that the eclipse is probably total. Knowing the
orbital period and the spectral type of the secondary (M4.5;
\citealt{SchwopeThomasBeuermann93}), we can use the predicted absolute
magnitude of the star (\citealt{Knigge06}; \citealt{KBP11}) to find a
distance of $290^{+110}_{-80}\,\mathrm{pc}$, in line with the
$\sim250\,\mathrm{pc}$ reported by
\cite{SchwopeThomasBeuermann93}. The smaller distance estimates of
\cite{Ciardi98} and \cite{SproatsHowellMason96} result from assuming a
later spectral type. \cite{Gansicke99} also estimated $\sim
180\,\mathrm{pc}$ from low-state UV data, but he notes that there was
still some accretion during the observation.

Finally, \cite{ReinschBurwitzBeuermann94} estimate a distance of
$300\pm60\,\mathrm{pc}$ for UW Pic, and \cite{Schwarz98} find
$620\pm100\,\mathrm{pc}$ for AI Tri, both also from the photometric
parallax of the donor.

\subsection{Systems for which only less reliable estimates are possible}
For the remaining 22 systems, we have to turn to a less direct method
of estimating distances. The donor star is expected to be responsible
for a large fraction of the near- and mid-IR flux of a CV. Therefore,
even in cases where the contribution of the donor star to a system's
IR light is not known, it is possible to use IR photometry to obtain
rough distance estimates. This is similar to the method of
\cite{Bailey81}, although we will use the predicted absolute donor
magnitudes of \cite{Knigge06} and \cite{KBP11}. Below, we will first
estimate the average contribution of the donor to the near- and mid-IR
flux of mCVs, and then use this to estimate distances for most of the
systems in our sample\footnote{\cite{Ak08} present another method of
  estimating distances from near-IR photometry, and also give
  distances for several of the mCVs in this sample.}.

\subsubsection{Finding the average mCV donor contribution to near- and mid-IR flux}
\cite{Knigge06} and \cite{KBP11} give the average differences between
the absolute $JHK$ magnitudes of CV donors predicted by their
semi-empirical sequence and the measured absolute magnitudes of a
sample of CVs (consisting mostly of non-magnetic systems, but
including also a few mCVs) with parallax distance estimates. This
gives an estimate of the typical donor contribution to the flux in
these bands, and, together with an apparent IR magnitude, can be used
to obtain a rough distance estimate (as well as a robust lower limit
on the distance).

In order to repeat this for mCVs alone, we use a sample of 23 magnetic
systems with orbital periods below 6 hours and reliable distance
estimates. The distances are based on trigonometric parallax (for 4
IPs and 5 polars), photometric parallax of the WD (5 polars), or
photometric parallax the donor (1 IP and 8 polars).  Not surprisingly,
we find that the typical donor contribution to the IR light is similar
for IPs and non-magnetic CVs (systems with discs), and smaller for
polars. Therefore, we group IPs and non-magnetic CVs together, and
consider the sample of polars separately. Furthermore, we decided to
omit the polars with published distances from photometric parallax of
the donor\footnote{Because it is possible that in these systems, the
  contribution to the IR light of, e.g., the accretion flow, is
  systematically low, given that spectral features of the donor stars
  are relatively prominent in the optical spectra.}; this means that
our average donor contributions for polars are based on a sample of
only 10 systems.

For the two samples of CVs with reliable distance estimates (IPs,
together with the non-magnetic CVs used by \citealt{Knigge06}, and the
sample of 10 polars) separately, we calculate the average of the
difference between the measured absolute magnitude of the CVs and the
absolute magnitude predicted for the donor star at that period by
\cite{Knigge06}. We do this for the 2MASS \citep{2mass} $JHK$ bands,
as well as the 2 shortest wavelength {\it WISE} \citep{wise} bands, W1
and W2. On average, the donor stars in polars emit, e.g., around 60\%
of the $K$-band light (compared to 31\%, for the sample of of mostly
non-magnetic CVs used by \citealt{Knigge06}). These results will be
presented in more detail in a future paper.

Clearly, a single number cannot describe the donor contribution to the
flux in some waveband for all systems and at all $P_{orb}$. However,
the sequence of \cite{Knigge06}, together with the typical fractions
of the near- and mid-IR flux contributed by the donor, should allow us
to predict absolute magnitudes to within the errors implied by the CV
samples used to find these fractions. We take the error in the
predicted absolute near- and mid-IR magnitudes to be the standard
deviation of the absolute magnitudes of these two samples from the
sequence, after it is offset to include the typical flux contribution
from sources other than the donor.

A final caveat is that the donor sequence may not be applicable to
long-period polars, because it is based on the masses and radii of
donor stars in non-magnetic CVs (which have MB above the orbital
period gap, while polars perhaps never experience MB). We are probably
safe in ignoring this difficulty, since we use the method for only 3
long-period polars, and since no single distance estimate strongly
affects our space density estimate\footnote{Furthermore, the high mass
  loss rates of non-magnetic CVs above the period gap cause the the
  donor radius to expand by at most $\simeq$30\%
  \citep{Patterson05,Knigge06}. The implied $\simeq$30\% distance
  error is too small to dominate the uncertainty in our estimates.}.

\subsubsection{Distance estimates relying on the estimated typical donor IR flux contribution}
\label{sec:CKdist}
We use the method outlined above to find distance estimates for 18 of
the mCVs in the RBS sample. In doing this, we assume that none of
these systems are period bouncers, or have evolved donor stars (except
for EI UMa, where we adopt an estimate from the literature that allows
for this; see below). We include the effect of interstellar extinction
in the distance estimates when using 2MASS photometry, but neglect it
when using {\it WISE} mid-IR photometry (extinction estimates for our
sources are discussed in Section~\ref{sec:xlum} below). We then find
the probability distribution function for the distance to each source,
assuming Gaussian errors in predicted absolute magnitudes, apparent
magnitudes, and extinction. The distance estimates listed in
Table~\ref{tab:distances} are the median, together with the 1-$\sigma$
confidence interval corresponding to the 16th and 84th percentile
points. We briefly discuss individual systems below.

CC Scl has DN outbursts (as several IPs do), but since it is certainly
not a typical DN (besides being an IP, its superoutbursts have
unusually low amplitude and short duration), we prefer not to base our
distance estimate on the outburst maximum here. We obtain a distance
of $200^{+110}_{-70}\,{\rm pc}$ from {\it WISE} photometry, slightly
smaller than the estimate of $360\pm 130\,{\rm pc}$ that
\cite{Patterson11} finds from the outburst maximum.

The distance of $330^{+180}_{-120}\,{\rm pc}$ that we find for AO Psc
using {\it WISE} photometry is consistent with the estimate of $\sim
250\,{\rm pc}$ obtained by \cite{Patterson84} from a relation between
the strength of emission lines and the absolute magnitude of the disc.

For EI UMa, we adopt the estimate of \cite{Reimer08}, which is based
on the same method we use here. Note that it is a tentative estimate
because this system has $P_{orb}$ in the range where the CV population
is expected to be dominated by systems with evolved donors (some
attempt was made to reflect this in the error; \citealt{Reimer08}).

EV UMa was in a high state when it was observed by 2MASS; we therefore
use $K=19.3(2)$ from \cite{OsborneBeardmoreWheatley94} to obtain
$d=690^{+270}_{-190}\,{\rm pc}$. \cite{OsborneBeardmoreWheatley94}
give a lower limit of $705\,\mathrm{pc}$ on the distance to this
system.

\cite{RamsayCropperMason04} find distances of $50$--$70\,\mathrm{pc}$
and $60$--$80\,\mathrm{pc}$ for GG Leo and EU UMa, respectively, from
UV data. They note, however, that these are likely underestimates,
since some of the UV flux might be from sources other than the WDs. We
find larger distances in both these cases (using {\it WISE} photometry
for GG Leo, and the 2MASS $H$ measurement for EU UMa).  Our estimate
of $d=170^{+90}_{-60}\,{\rm pc}$ for GG Leo is consistent with limits
of $d>100\,\mathrm{pc}$ (based on the optical spectrum;
\citealt{Burwitz98}) and $d<300\,\mathrm{pc}$ (from the requirement
that the implied accretion rate is not unreasonably high;
\citealt{Brinkworth07}).

We use {\it WISE} data of V393 Pav to estimate
$d=340^{+170}_{-110}$~pc, in agreement with the estimate of $d \sim
350\,\mathrm{pc}$ by \cite{ThomasBeuermannSchwope96}.

V4738 Sgr and RS Cae are not detected by 2MASS or {\it WISE}; we find
$H=17.61(9)$ and $18.1(1)$, for these two systems, from images
obtained with the Infrared Survey Facility (IRSF; see
e.g.\ \citealt{GlassNagata00} and \citealt{Nagayama03}) at the South
African Astronomical Observatory. This gives distance estimates of
$250^{+90}_{-70}\,{\rm pc}$ and $880^{+330}_{-240}\,{\rm pc}$ for
V4738 Sgr and RS Cae, respectively.  Note that although the radial
velocity curve of RS Cae is aliased \citep{BurwitzReinschSchwope96}
the uncertainty in period does not significantly affect the distance
estimate.  From the absence of M star features in the optical spectra,
lower limits of $d>440\,\mathrm{pc}$ for RS Cae
\citep{BurwitzReinschSchwope96} and $d>190$~pc for V4738 Sgr
\citep{BurwitzReinschBeuermann97} have been derived.

\cite{ClaytonOsborne94} report $K=16.4(7)$ for EK UMa (which is also
not detected by 2MASS or {\it WISE}), and give a lower limit of
$d>410\,\mathrm{pc}$. We use their $K$-band magnitude to estimate
$d=590^{+340}_{-210}\,\mathrm{pc}$ for this system.

\cite{Patterson84} estimates $d \sim 400\,\mathrm{pc}$ for AN UMa
(\citealt{bible} lists a limit of $>270$~pc; see also
\citealt{Liebert82} for some discussion on lower distance limits). We
find $300^{+150}_{-100}\,\mathrm{pc}$, using {\it WISE} photometry.

For V1007 Her, we use the W1 band of {\it WISE} to find a distance of
$560^{+280}_{-180}\,\mathrm{pc}$. \cite{GreinerSchwarzWenzel98} give a
lower limit of $d>250\,\mathrm{pc}$.

We obtain a distance of $500^{+250}_{-170}\,\mathrm{pc}$ for CW Hyi,
again from the W1-band magnitude. Our estimate is slightly large
compared to the distance of $\sim 250$~pc reported by
\cite{SchwopeBrunnerBuckley02}; however, the donor star features are
not prominent in the optical spectrum, implying that the estimate of
the donor contribution to the optical flux is quite rough
\citep{SchwopeBrunnerBuckley02}.

Our distance estimate of $380^{+190}_{-120}\,\mathrm{pc}$ for RX
J1610.1+0352 (also called RBS 1563) from {\it WISE} photometry is in
good agreement the estimate of 320~pc that \cite{Ak08} obtain from
2MASS data.

We use the 2MASS $H$-band magnitude of 1RXS J231603.9-052713 (RBS
1973) to find a distance of
$460^{+170}_{-120}\,\mathrm{pc}$. \cite{Rodrigues06} have estimated
$\simeq 410\,\mathrm{pc}$, in agreement with our value, from a model
of the optical and infrared flux that includes cyclotron emission and
a heated secondary.

There is no published information on the distances of EQ Cet, EG Lyn,
and RX J0859.1+0537 (RBS 734). We use {\it WISE} detections of these 3
systems in our distance estimates.

\subsubsection{Systems for which we obtain only very weak distance constraints}
\label{sec:problems}
We are now left with 4 systems for which estimating distances presents
more serious problems. These are CD Ind, IW Eri, CV Hyi, and FH
UMa. For CD Ind and IW Eri, the method discussed above produces
distance estimates that are inconsistent with other information, while
no near- or mid-IR detections are available for CV Hyi and FH UMa. We
discuss these systems in turn below. In Section~\ref{sec:sensdistance}
we will show that these four very poorly constrained distances do not
have an important effect on our space density estimate. This is
because we are able to place lower limits on the distances, and, in
all 4 cases, the resulting upper limit on the contribution of these
systems to the total space density is small.

Apparent magnitudes in the 2MASS and {\it WISE} bands, together with
our assumptions about typical donor flux contributions in the IR
bands, would imply a distance of roughly $110\,\mathrm{pc}$ for CD
Ind. However, \cite{SchwopeBuckleyODonoghue97} find no sign of the
donor in their optical spectra, and give a lower limit of of $d>250$,
assuming a donor spectral type of M5.

For IW Eri, {\it WISE} magnitudes would yield a distance of only about
80~pc, which is inconsistent with its non-detection in 2MASS. We
measure $K>15.5$ (from the faintest detections in the area around IW
Eri), implying $d>150\,\mathrm{pc}$.

Both CD Ind and IW Eri have low- and high photometric states
(\citealt{SchwopeBuckleyODonoghue97};
\citealt{SchwopeBrunnerBuckley02}). It is possible that CD Ind was
observed in a high state by 2MASS as well as {\it WISE}, while IW Eri
was caught in a high state by {\it WISE} and in a low state by
2MASS. The high-state optical spectrum of CD Ind shows rising flux
towards the red end, interpreted as cyclotron emission caused by a
weak magnetic field ($B\simeq 11\,\mathrm{MG}$;
\citealt{SchwopeBuckleyODonoghue97}). Therefore, CD Ind likely has
unusually large cyclotron emission in the IR, because of its unusually
low magnetic field strength.

For CV Hyi and FH UMa, we find lower limits of $d>300\,\mathrm{pc}$
and $d>320\,\mathrm{pc}$, respectively, from {\it WISE}
non-detections. \cite{BurwitzReinschBeuermann97} also report
$d>300\,\mathrm{pc}$ for CV Hyi, based on its optical spectrum.

We can derive conservative upper limits on the distances of these 4
systems from what would be implausible optical luminosities, or even
position in the galaxy. However, since such limits are very weak, they
are not useful here.  It does seem that IW Eri is not very distant,
since it has a proper motion of about $45\,\mathrm{mas/yr}$
(\citealt{Monet03}; \citealt{Petersthesis}).  \cite{Patterson11} finds
a tangential velocity of $39 \pm 5\,\mathrm{km/s}$ for a sample of
normal non-magnetic CVs. If we assume that IW Eri comes from a
population with the same age, we can tentatively estimate a distance
of $\sim 180\,\mathrm{pc}$.

For these 4 systems we will simply assume distance distributions that
are Gaussian in $\log (d)$, with $\sigma_{log(d)}$ chosen so that the
16th percentile corresponds to the lower limits given above. Of course
this is not correct, but our space density estimate is not strongly
affected (see Section~\ref{sec:sensdistance}).

\subsection{Possible bias in distance estimates}

Distance estimates used here may suffer from the well-known Malmquist
(\citealt{Mbias}; in the case of photometric parallax) and Lutz-Kelker
(\citealt{LKbias}; for trigonometric parallax) biases. We are not
concerned about Lutz-Kelker bias in the parallax measurements of EX
Hya, TV Col and EF Eri, because the parallax errors are very small in
the first two cases (\citealt{McArthur01}; \citealt{Beuermann03}), and
because the bias was considered by \cite{Thorstensen03} in the case of
EF Eri. This leaves the possibility of Malmquist bias in the
photometric parallax distances presented in
Section~\ref{sec:photpardist} and \ref{sec:CKdist}. We have checked
how large this effect could be, in the same way as described in
\cite{nonmagphi}. We find that the bias, if present at all (see the
more detailed discussion in \citealt{nonmagphi}), is in each case
insignificant.

\section{X-ray spectra and luminosities}
\label{sec:xlum}

Polars are luminous soft X-ray sources, and a large number was
detected by \emph{ROSAT} (e.g.\ \citealt{BeuermannThomasReinsch99};
\citealt{Thomas98}). Their X-ray spectra are usually modelled as
blackbodies, with $kT\sim 20$ to $40\,\mathrm{eV}$
(e.g.\ \citealt{RamsayCropperMason96,ThomasBeuermann98}; Table 6.5 of
\citealt{bible}).  Although a harder component is often present, in
most polars it only contributes a significant part of the flux in the
\emph{ROSAT} band during low states \citep{BeuermannThomas93}, where
even very nearby polars are too faint to be included in the RBS (see
also Section~\ref{sec:limits}). A handful of confirmed and candidate
polars that show no soft X-ray component, even in the high state, is
now known
(e.g.\ \citealt{RamsayCropper04,RamsayCropper07,Ramsay09,Vogel08}). However,
these systems do not appear to be intrinsically very common. Only 2
are within 200~pc, and e.g.\ \emph{XMM-Newton} and \emph{INTEGRAL} and
have not discovered many new polars (despite the nearest of the known
hard polars, namely BY Cam, V1432 Aql, and V2301 Oph, being detected
in the \emph{INTEGRAL/IBIS} survey).

IPs have hard X-ray spectra, often modelled as intrinsically absorbed
thermal bremsstrahlung with $kT \sim 30\,\mathrm{keV}$
(e.g.\ \citealt{Patterson94}; \citealt{FKR85}). In a few IPs, a soft
component is present (e.g.\ \citealt{HaberlMotch95};
\citealt{deMartino04}), but our sample does not contain any of these
``soft IPs''.  Depending on the accretion geometry, IPs can have very
high intrinsic absorption (the intrinsic absorption also varies in
individual systems). We take the intrinsic $N_H$ as $2 \times
10^{20}\,\mathrm{cm^{-2}}$ for EX Hya and DO Dra (\citealt{Richman96};
\citealt{Mukai03}), and $4 \times 10^{22}\,\mathrm{cm^{-2}}$ for CC
Scl, AO Psc, TV Col, and EI UMa (\citealt{Pietsch87};
\citealt{Rana04}; \citealt{Ramsay08}; \citealt{Woudt12}), with a
covering fraction of 0.5 in all cases.

We assume a $kT=30\,\mathrm{eV}$ blackbody spectrum for the polars,
and a $kT=30\,\mathrm{keV}$ thermal bremsstrahlung spectrum for the
IPs in our sample. These single temperature, single component spectra
are not very physical (e.g.\ \citealt{NortonWatson89};
\citealt{Mukai03}; \citealt{BeuermannBurwitzReinsch12}), and they are
not expected to be good approximations for all sources in the
sample. The assumed intrinsic absorption for the IPs in the sample are
also rough approximations. Therefore, the values we give for $L_X$
should be treated with caution.  However, the space density estimate
is much less sensitive to both the assumed X-ray spectra and the
amount of intrinsic absorption we adopt for the IPs than are the
estimated $L_X$ values (we will return to this in
Section~\ref{sec:sensspec}).

Since this is a high Galactic Latitude sample, interstellar absorption
is low for all our sources. A few systems are at sufficiently large
distances that the total Galactic $N_H$ (as given by
\citealt{Kalberla05}) is a good approximation. Many of the polars have
$N_H$ estimates from X-ray spectral fits
(e.g.\ \citealt{SchwopeBuckleyODonoghue97,ReinschBurwitzBeuermann94,Schwarz09,BarrettSinghMitchell99,RamsayCropperMason96,Ramsay94,PandelCordova05,Schwope07,BeuermannThomasPietsch91,Burwitz98,Singh95,GreinerSchwarzWenzel98,BurwitzReinschBeuermann97}). For
several systems, there are absorption measurements based on UV data
(e.g.\ \citealt{Araujo-BetancorGansickeLong05,LaDous91,Verbunt87}),
and for a few more, \cite{BruchEngel94} give $A_V$ estimates (we use
the dust to gas ratio of \citealt{PredehlSchmitt95} to obtain
$N_H$). In cases where we have no observational estimate of
interstellar absorption, we use the models of \cite{AmoresLepine05}
and \cite{Drimmel03}.

We assume 50\% errors in estimates of interstellar column densities,
and give unabsorbed X-ray luminosities in the 0.12--2.48~keV band in
Table~\ref{tab:distances}.

\section{Calculating the space density}
\label{sec:rhocalc}

\subsection{The method}
\label{sec:calc}
We describe the calculation that gives the space density, together
with the uncertainty on that estimate only briefly. More detailed
discussions may be found in \cite{nonmagphi} and \cite{NEPrho}.

The effective observed volume of the RBS depends on the spacial
distribution of CVs in the Galaxy, and on the X-ray luminosities of
systems in the sample (since the survey is flux limited, rather than
volume limited). It is found using the relation given by
e.g.\ \cite{StobieIshidaPeacock89} and \cite{TinneyReidMould93}:
$$V_j=\Omega \frac{h^3}{|\sin
  b|^3}\left[2-\left(x_j^2+2x_j+2\right)\mathrm{e}^{-x_j}\right].$$
The index $j$ represents mCVs in the sample; $\Omega$ is the solid
angle observed in the survey\footnote{Because $b$ is variable over
  $\Omega$, we compute each $V_j$ as a sum over smaller solid angles,
  $\delta\Omega$. The terms in this sum are slices subtended by
  $2^\circ$ in $b$. These $\delta\Omega$ are sufficiently small that
  the error introduced by $b$ varying over $\delta\Omega$ is
  negligible.}, and $x_j=d_j |\sin b|/h$, with $d_j$ the maximum
distance at which system $j$ could have been detected (a function of
its luminosity, the flux limit, and extinction along all lines of
sight covered by the survey). This relation assumes that $\rho$ does
not vary with radial position in the Galaxy, and that the vertical
density profile is exponential. We assume scale-heights of
$h=260\,\mathrm{pc}$ for short-period systems, and
$h=120\,\mathrm{pc}$ for long-period systems\footnote{The reason for
  this is that short- and long-period CVs are expected to be old and
  young populations, respectively. It is only a crude approximation,
  since systems can form at short period, and since polars at all
  periods are believed to evolve slowly. Our results do not change
  significantly if we simply assume the same $h$ for all systems (see
  Section~\ref{sec:sensh}.)}. To find $d_j$, we need $N_H$ as a
function of distance and $b$. For this we assume an exponential
vertical density profile for gas, with a scale-height of
$140\,\mathrm{pc}$, and normalized to give $N_H$ corresponding to
$A_V=0.8\,\mathrm{mag}/\mathrm{kpc}$ in the Galactic
Plane\footnote{Since the survey covers a large area, at high galactic
  latitudes (all $|b|>30^\circ$), and the maximum distances we deal
  with are large compared to the typical size of inhomogeneities in
  the ISM, this smooth model for the density of gas is justified (for
  the most part, but see Section~\ref{sec:limits}). We explore the
  sensitivity of our $\rho$ estimate to the assumed mid-Plane gas
  density in Section~\ref{sec:completeness}.}.

The sum of the contributions of the 30 systems in the sample then
gives the mid-Plane value of $\rho$ ($\rho_0=\sum_j 1/V_j$). For the
moment ignoring all errors, and simply assuming that the best-estimate
distance, count rate, and $N_H$ are the true values, we list the
fractional contribution that each system in our sample makes to the
total space density ($\rho_j/\rho_0$) in the 8th column of
Table~\ref{tab:distances}. This shows that our $\rho$ measurement is
not dominated by 1 or 2 objects.

In order to correctly sample the full parameter space allowed by the
data, and hence find the error on $\rho_0$, we compute its probability
distribution function using a Monte Carlo simulation that calculates
$\rho_0$ for a large number of mock samples. The mock samples are
created by drawing a distance, $N_H$, and count rate for every
observed system from the appropriate distribution. These values are
used to calculate each $1/V_j$, which is then weighted by a factor
$\mu_j$, drawn from the probability distribution of the number of
sources belonging to the population (corresponding to a particular
observed system) that one expects to detect in the RBS (see
\citealt{NEPrho} and \citealt{nonmagphi}).

\subsection{Results}
\label{sec:results}
We have carried out this calculation for the whole observed sample of
30 mCVs, as well as for several sub-samples (polars, IPs, short- and
long-period systems, and short-period polars and long-period
IPs). Table~\ref{tab:res} summarizes the results.

\begin{table}
 \centering
  \caption{Space density estimates for all mCVs, IPs and polars separately, short- and long-period mCVs, and short-period polars and long-period IPs. UW Pic and J$0859+05$ are excluded from the samples split by $P_{orb}$, because their orbital periods fall in the period gap. In addition to the values inferred from the observed RBS sample, we list the estimates that would be obtained if polars are undercounted by a factor of 2, because of a high-state duty cycle of only 0.5.}
  \label{tab:res}
%
  \begin{tabular}{@{}lcc@{}}
  \hline
Sample                & \multicolumn{2}{c} {$\rho_0/(10^{-7}\,\mathrm{pc^{-3}})$}  \\
                      & No assumed low states    & Polar duty cycle 0.5 \\
 \hline
All mCVs              & $8.2^{+3.9}_{-2.3}$ & $13^{+6.3}_{-3.8}$ \\
Polars                & $4.9^{+2.7}_{-1.5}$ & $9.8^{+5.4}_{-3.1}$ \\
IPs                   & $2.7^{+2.4}_{-1.2}$ &  \\
short-$P_{orb}$ mCVs   & $4.6^{+3.2}_{-1.7}$ & $8.3^{+5.7}_{-3.0}$ \\
long-$P_{orb}$ mCVs    & $2.8^{+1.8}_{-1.1}$ & $3.8^{+2.2}_{-1.4}$ \\
short-$P_{orb}$ polars & $3.6^{+2.5}_{-1.9}$ & $7.2^{+5.0}_{-2.7}$ \\
long-$P_{orb}$ IPs     & $1.7^{+1.6}_{-0.9}$ &  \\
 \hline
 \end{tabular}
\end{table}

\subsubsection{The probability distribution function of $\rho_0$}
\label{sec:rhoresult}

\begin{figure}
 \includegraphics[width=84mm]{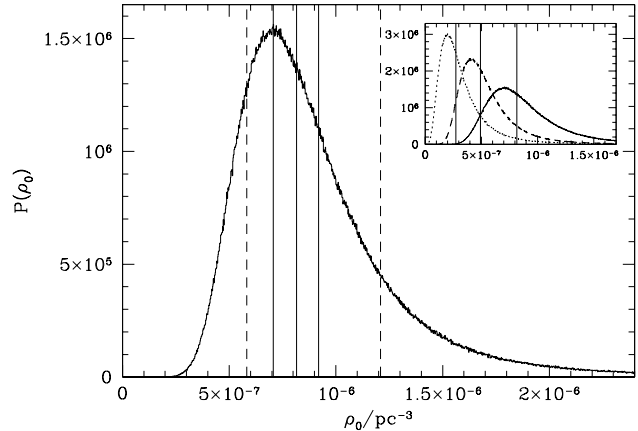} 
  \caption {The $\rho_0$ distribution for all mCVs, resulting from our simulation. Solid lines mark the mode, median, and mean at $7.1 \times 10^{-7}$, $8.2 \times 10^{-7}$, and $9.2 \times 10^{-7}\,\mathrm{pc^{-3}}$. Dashed lines show a 1-$\sigma$ interval from $5.8 \times 10^{-7}$ to $1.2 \times 10^{-6}\,\mathrm{pc^{-3}}$. The probability distribution functions shown in the inset are (with modes from high to low $\rho_0$) for the whole mCV sample, polars alone (dashed), and IPs alone (dotted). In the inset, solid lines at $8.2 \times 10^{-7}$, $4.9 \times 10^{-7}$, and $2.7 \times 10^{-7}\,\mathrm{pc^{-3}}$ mark the medians of the three distributions.
}
 \label{fig:rhohist}
\end{figure}

The distribution of $\rho_0$ values, normalized to give a probability
distribution function, from the simulation including all magnetic
systems is shown in Fig.~\ref{fig:rhohist}. The mode, median, and mean
of the distribution are marked by solid lines at $7.1 \times 10^{-7}$,
$8.2 \times 10^{-7}$, and $9.2 \times 10^{-7}\,\mathrm{pc^{-3}}$,
while the dashed lines at $5.8 \times 10^{-7}$ and $1.2 \times
10^{-6}\,\mathrm{pc^{-3}}$ show a 1-$\sigma$ confidence interval (the
16th and 84th percentile points of the distribution). In other words,
our estimate for the mid-plane space density of mCVs is $8^{+4}_{-2}
\times 10^{-7}\,\mathrm{pc^{-3}}$. We find that the large errors in
many of the distances dominate the total uncertainty in $\rho_0$,
although the small sample size and $N_H$ errors also contribute
significantly.

The inset in Fig.~\ref{fig:rhohist} shows the probability distribution
functions of $\rho_0$ for the whole mCV sample again, as well as for
IPs (dotted histogram with lowest mode) and polars (dashed histogram)
separately. We find space densities of $3^{+2}_{-1} \times
10^{-7}\,\mathrm{pc^{-3}}$ for IPs and $5^{+3}_{-2} \times
10^{-7}\,\mathrm{pc^{-3}}$ for polars.

\subsubsection{Upper limits on the space density of an undetected population}
\label{sec:limits}

The estimate above assumes that the detected population of mCVs is
representative of the true underlying population, in the sense that it
contains at least 1 of the faintest IPs and polars that occur in the
intrinsic population (but not that it contains faint and bright
systems in proportion to their intrinsic incidence; see
\citealt{nonmagphi}). Since the effective volume of the survey is
smaller for fainter $L_X$, it is possible that even a large population
of sources at the faint end of the luminosity function can go
completely undetected.  Here we place limits on the sizes of faint
populations of polars and IPs that could escape detection in the RBS.

We again perform a Monte Carlo simulation with the same simple Galaxy
model for stars and gas as described in Section~\ref{sec:calc} above
(but, for simplicity, now assuming a single scale height of $260$~pc
for all mCVs). A model population of mCVs, all with the same $L_X$,
and with a spectrum appropriate to either polars or IPs\footnote{For
  polars, we again assume a $kT=30\,\mathrm{eV}$ blackbody spectrum,
  and for IPs, we assume a $kT=30\,\mathrm{keV}$ thermal
  bremsstrahlung spectrum with a partial covering absorber with
  covering fraction of 0.5 and $N_H=2 \times
  10^{20}\,\mathrm{cm^{-2}}$.}, is distributed in the model Galaxy, in
order to find the value of $\rho_0$ for which the predicted number of
detected systems is 3 (so that detecting 0 such systems is a
2-$\sigma$ result). We do this for a range of $L_X$ below the faintest
values found for the observed sample.

\begin{figure}
 \includegraphics[width=84mm]{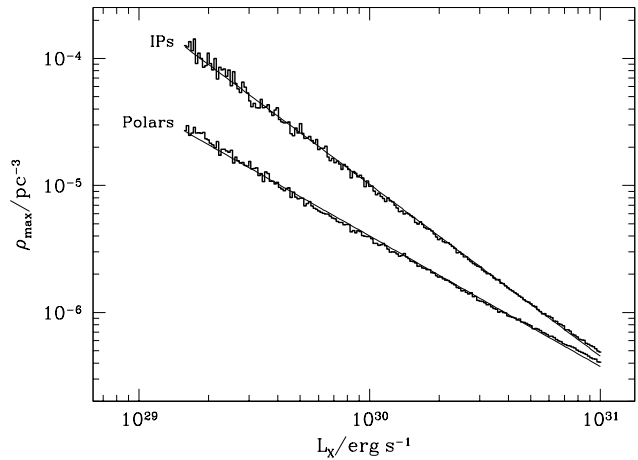} 
  \caption {The upper limit on the mid-plane space density as a function of X-ray luminosity for an undetected population of polars and IPs. The data from the simulations are shown as a bold histogram, and fits are over-plotted as a finer lines. Note that the assumed X-ray spectra of polars and IPs are different, hence the different slopes.
}
 \label{fig:limit}
\end{figure}

Fig.~\ref{fig:limit} shows the maximum allowed $\rho_0$ as a function
of $L_X$, separately for possible undetected polar and IP
populations. The limits from the simulation are plotted as bold
histograms, and the fine curves are fits to the data, given by
$$\rho_{max}= 1.02\times 10^{-5} (L_X/10^{30}\,\mathrm{erg\,s^{-1}})^{-1.35}\,\mathrm{pc^{-3}}$$ 
for IPs, and 
$$\rho_{max}= 4.01\times 10^{-6} (L_X/10^{30}\,\mathrm{erg\,s^{-1}})^{-1.03}\,\mathrm{pc^{-3}}$$ 
for polars.  $\rho_{max}$ is only the upper limit on the mid-plane space
density of an undetected population, and does not include the
contribution from the observed systems. Thus, as a specific example, a
population of undetected polars with a space density as high as $5
\times$ the measured $\rho_{polar}$ must have $L_X \la 10^{30}\,{\rm
  erg s^{-1}}$. A hidden population of IPs can only have $\rho_0 = 5
\times \rho_{IP}$ if it consists of systems with X-ray luminosities
fainter than $5 \times 10^{30}\,{\rm erg s^{-1}}$.

Because the two relations given above involve $L_X$, which depends
quite sensitively on our assumed X-ray spectra\footnote{For example,
  adopting blackbody temperatures of 10 and 50~eV for a moderately
  absorbed polar ($N_H=2\times 10^{20}\,\mathrm{cm^{-2}}$), yield
  values of $L_X$ that differ by a factor of 20. For the hard spectra
  of IPs, $L_X$ is less sensitive to the assumed temperature of the
  bremsstrahlung spectrum, but does vary widely over an intrinsic
  $N_H$ range as large as is observed.}, they are of only limited use
(this is discussed further in Section~\ref{sec:sensspec}).

This is also an instance where our smooth model of the ISM is not a
valid approximation. This is because, e.g., polars with $L_X =
10^{29}\,{\rm erg s^{-1}}$ can only be detected out to $\simeq
40\,{\rm pc}$, which places them inside the `Local Bubble'
(e.g.\ \citealt{FrischRedfieldSlavin11}). However, since our model
probably gives too high absorption at small distances, it only means
that, at least for polars, the upper limit on $\rho_0$ is conservative
at the faint $L_X$ end. In the case of IPs, the structure of the local
ISM matters less, since their spectra are less affected by
interstellar absorption.

Polars probably do not experience a large range in secular $\dot{M}$
over the course of their evolution (because $\dot{M}$ is a relatively
flat function of $P_{orb}$ when the only AML mechanism is GR). This
means that we do not expect the polar luminosity function to rise
towards the faint $L_X$ end, beyond the luminosities that we are
sensitive to. However, most (possibly all) polars switch between low-
and high $\dot{M}$ states on shorter timescales (e.g.,
\citealt{Ramsay04}). In the low photometric state, even very nearby
polars are too faint to be included in the RBS. An example is AR
UMa. It is at a distance of only 86~pc
\citep{ThorstensenLepineShara08}, and is a bright soft X-ray source in
its high state (e.g.\ \citealt{Remillard94}), but was so faint during
the RASS that it is not even included in the faint source
catalogue. The deep low states of polars imply that in a single-epoch
X-ray survey (which is what the RASS was over most of the sky) only a
fraction of the local polar population that is equal to the high-state
duty cycle can be detected. In other words, our assumption that the
polar sample detected in the RBS is representative of the intrinsic
population, is not strictly valid, since it does not include low-state
systems. \cite{Ramsay04} find that polars spend roughly half their
time in low states, implying that our estimate of $\rho_{polar}$ is
probably a factor of 2 too low\footnote{Note that the high-state duty
  cycle of polars is not very well constrained; see 
  e.g.\ \cite{Hessman00}, \cite{Araujo-BetancorGansickeLong05},
  \cite{WuKiss08}, and \cite{Breedt12}.}. In the final column of
Table~\ref{tab:res}, we list the space density estimates that we
obtain when allowing for a high-state duty cycle of 0.5 for polars and
assuming that polars are undetectable their low states.

\subsection{Sensitivity of the results to uncertain assumptions}
\label{sec:sens}
Here we examine the sensitivity of our space density estimates to some
of the more uncertain assumptions we made in calculating it. We will
consider the possible impact of the systems with the most poorly
constrained distances, the scale-heights we adopt for different CV
populations, the assumed amount of interstellar absorption, the
completeness of the sample at lower $|b|$, and the assumed X-ray
spectra. By investigating an extreme range of assumptions, we show
that possible systematic errors affect these space density estimates
to less than roughly a factor of 2.

\subsubsection{The most uncertain distance estimates}
\label{sec:sensdistance}
Four members of the mCV sample, CV Hyi, FH UMa, IW Eri, and CD Ind,
have very poorly constrained distances (see
Section~\ref{sec:problems}). The lower limits on distances are,
however, always robust, and allow us to determine that these systems
make at most a small contribution to the overall space
density. Specifically, if we assume in each case the smallest allowed
distance, the median of the $\rho_0$ distribution would be at
$8.7\times 10^{-7}\,\mathrm{pc^{-3}}$, while neglecting the
contributions of these 4 systems yield $6.6\times
10^{-7}\,\mathrm{pc^{-3}}$. This is within the errors of our best
estimate of $8^{+4}_{-2}\times 10^{-7}\,\mathrm{pc^{-3}}$. At its
minimum allowed distance, IW Eri would account for 0.1 of the total
space density, while the other 3 would all contribute less.

\subsubsection{The scale height of mCVs}
\label{sec:sensh}
We have assumed scale-heights of $260$ and $120\,\mathrm{pc}$ for
short- and long-period systems, respectively, as a crude approximation
to the expectation that short- and long-period CVs belong to
populations of different typical ages. If we instead set $h$ to $260$
or $120\,\mathrm{pc}$ for all systems, we obtain $\rho_0$
distributions with medians at $6.3\times 10^{-7}\,\mathrm{pc^{-3}}$
and $14\times 10^{-7}\,\mathrm{pc^{-3}}$, respectively. Therefore,
although a plausible range of scale-heights shift the $\rho_0$
distribution considerably, one would obtain estimates consistent with
our result for other reasonable assumptions regarding the Galactic
distribution of mCVs.

We have also checked that the $z$-distribution of the RBS mCV sample
is consistent with the Galaxy model we use. In this calculation, we
use the same model for the density of stars and gas, and assume the
same X-ray spectra as before\footnote{Since the RBS is a high Galactic
  latitude, flux-limited survey, the observed sample does not have the
  same $z$-distribution as the underlying population. Generating a
  model $z$-distribution therefore involves an assumption regarding
  the luminosity function of the intrinsic polar and IP
  populations. We have assumed intrinsic distributions that are
  Gaussian in $\log(L_X)$, and experimented with several values for
  the average and standard deviation.}. After imposing the RBS flux-
and $|b|$-limits, we use a Kolmogorov-Smirnov (KS) test to compare
model and observed $z$-distribution. We find that the whole sample of
30 observed systems, as well as subsamples consisting of short- and
long-period systems, have $z$-distributions that are consistent with
both $h=260$ and $120\,\mathrm{pc}$. In other words, our sample is too
small, and the distance errors are too large, to distinguish between
these two choices of scale height.

\subsubsection{Interstellar absorption and completeness of the sample}
\label{sec:completeness}
Our treatment of interstellar absorption is very simple, and could be
a concern, considering the soft, easily absorbed spectra of polars. As
noted before, the unrealistically smooth model of the interstellar
medium is not an important shortcoming (since we are using it to find
a survey volume which covers a large fraction of the whole sky, at
relatively high $|b|$, and since even the smaller maximum distances
involved in the calculation are large enough to implying that we can
average over regions of low and high absorption). The total amount of
absorption, on the other hand, could have a large effect on the space
density estimate, because higher absorption (along all lines of sight)
reduces the survey volume.

We have set the mid-plane density of gas to produce
$A_V=0.8\,\mathrm{mag}/\mathrm{kpc}$ for $b=0^\circ$. A wide range in
average mid-plane extinction has been reported. \cite{Drimmel03} give
$A_V \simeq 0.7\,\mathrm{mag}/\mathrm{kpc}$ for lines of sight near
the Galactic Plane in the inner disc of the Galaxy, and $A_V \simeq
0.5\,\mathrm{mag}/\mathrm{kpc}$ in the outer Galactic disc. The model
of \cite{AmoresLepine05} produces $A_V \simeq
1\,\mathrm{mag}/\mathrm{kpc}$, and \cite{Vergely98} find
$1.2\,\mathrm{mag}/\mathrm{kpc}$ in the Galactic Plane. Repeating our
space density calculation with the density of interstellar gas
normalized to give values of 0.5 to $1.2\,\mathrm{mag}/\mathrm{kpc}$
in the Plane, yields $\rho_0$ distributions with median values of
$5.8\times 10^{-7}$ and $12\times 10^{-7}\,\mathrm{pc^{-3}}$,
respectively\footnote{$\rho_{IP}$ on its own is much less sensitive to
  this, because of the harder spectra of IPs, and because intrinsic
  absorption reduces the importance of interstellar absorption.}. This
is consistent with our best estimate of $8^{+4}_{-2}\times
10^{-7}\,\mathrm{pc^{-3}}$.

We can also check for evidence that the completeness of our sample
decreases at lower $|b|$, as one might expect from increasing
absorption along lines of sight closer to the Galactic Plane. To do
this we simply create subsamples with different $|b|$
cutoffs. Compared to $\rho=8^{+4}_{-2}\times
10^{-7}\,\mathrm{pc^{-3}}$ from the whole sample of 30 systems
($|b|>30^\circ$), we find median $\rho_0$ values of $8.1\times
10^{-7}\,\mathrm{pc^{-3}}$ and $5.6\times 10^{-7}\,\mathrm{pc^{-3}}$
for $|b|>40^\circ$ (17 systems) and $|b|>50^\circ$ (10 systems),
respectively. These number are consistent to within the errors, and if
there is a trend, it goes in the opposite direction than expected if
the sample is less complete at lower $|b|$.

\subsubsection{The assumed X-ray spectra}
\label{sec:sensspec}
A final source of systematic uncertainty associated with the space
density estimate is our assumed X-ray spectra. We have used simple,
single component spectra for all systems (bremsstrahlung spectrum for
IPs, and blackbodies polars). Here we will check the possible impact
of a range of X-ray spectra on the space density estimates.

The space density calculation uses the maximum distance at which a
given mCV in the observed sample could have been detected, since this
determines the survey volume (for a population of systems with the
same $L_X$). For a given object, the maximum distance depends on the
ratio of $F_X$ to the flux limit, which may as well be expressed as a
ratio of observed count rate to limiting count rate. This means that
the assumed X-ray spectral shape has relatively little influence on
the space density calculation (only the interstellar absorption
depends on the X-ray spectrum, and this only matters to the extent
that $N_H$ differs between the true and maximum distance, and that the
slope of absorption as a function of $N_H$ differs for different
spectra).

\begin{figure}
 \includegraphics[width=84mm]{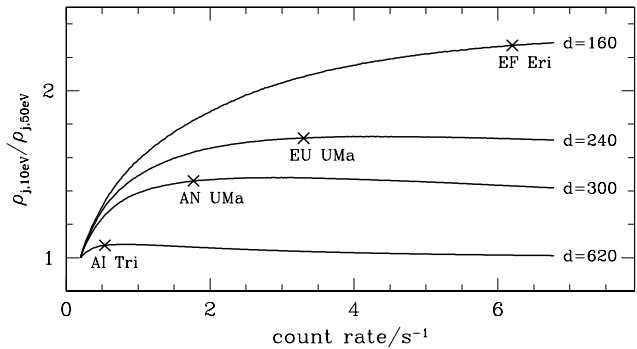} 
  \caption {The ratio of the values of $1/V_j$ obtained by assuming $kT=10$ and $50\,\mathrm{eV}$ blackbody spectra, as a function of the count rate at which the system is detected. The 4 curves are for different true distances. The count rates of 4 of the polars in our sample are indicated on the curves corresponding to their best-estimate distances. The largest effect of different spectra on $\rho_j$ occurs for a system that lies at a small distance compared to the maximum distance it could have been detected at (since this implies a large amount of absorption between $d$ and $d_{max}$). The most extreme example in the RBS is EF Eri, where its contribution to the total space density would be a factor of about 2.2 larger for an assumed $kT=10\,\mathrm{eV}$ than for an assumed $kT=50\,\mathrm{eV}$ spectrum. 
}
 \label{fig:rhosens}
\end{figure}

To show this more explicitly, if $y$ is the ratio of count rate ($c$)
to X-ray flux (i.e. $y=c/F_X$, a function of the spectrum, the
instrument response, and $N_H$), then the distance is
$d=\sqrt{yL_X/4\pi c}$, so $d_{max}=\sqrt{y_{max}c/yc_{lim}}$, where
$y_{max}=y(d_{max})$ and $c_{lim}$ is the limiting count rate. Then,
for $d\simeq d_{max}$ (the case of a system detected at close to the
limiting count rate), and for large $d$ ($N_H$ at $d$ already at about
the total value for the Galaxy), $y_{max} \simeq y$, implying that the
shape of $y(N_H)$ does not matter, and different spectra give the same
$d_{max}$ (and therefore $1/V_{max}$). Otherwise, the difference in
$d_{max}$ resulting from assuming different spectra depends on the
difference in slope of $y(N_H)$ for those spectra. We illustrate this
in Fig.~\ref{fig:rhosens}, where we use blackbody spectra with $kT=10$
and $50\,\mathrm{eV}$.  The figure shows that, while different assumed
spectra imply very different values of $L_X$, the space density
contribution of a given system changes by at most a factor of $\simeq
2$, with a much smaller effect in most cases. The effect is also
smaller for spectra appropriate to IPs than for the soft spectra of
polars.

For the whole sample of polars, we would obtain median $\rho_0$ values
of $8.1\times 10^{-7}\,\mathrm{pc^{-3}}$ and $4.5\times
10^{-7}\,\mathrm{pc^{-3}}$ assuming a $kT=10$ and $50\,\mathrm{eV}$
blackbody spectrum, respectively, compared to
$\rho_{polar}=5^{+3}_{-2}\times 10^{-7}\,\mathrm{pc^{-3}}$ for our
favoured $30\,\mathrm{eV}$ spectrum. Note that while a
$kT=30\,\mathrm{eV}$ blackbody spectrum is unlikely to be a good
approximation for all the polars in our sample, we also do not expect
all to be better represented by either much lower or much higher
temperature blackbodies.

$\rho_{IP}$ is less sensitive to the assumed spectrum, because the
difference in absorption between different hard, intrinsically
absorbed spectra is small, even for a large range in assumed
temperature. Our best model is a $kT=30\,\mathrm{keV}$ thermal
bremsstrahlung spectrum, with intrinsic column densities of $2 \times
10^{20}\,\mathrm{cm^{-2}}$ for EX Hya and DO Dra, and $4 \times
10^{22}\,\mathrm{cm^{-2}}$ for CC Scl, AO Psc, TV Col, and EI UMa, all
with a covering fraction of 0.5. We will separately consider different
bremsstrahlung temperatures and intrinsic absorption. With $kT=10$ and
$50\,\mathrm{keV}$, respectively, we obtain median $\rho_0$ values of
$2.8\times 10^{-7}\,\mathrm{pc^{-3}}$ and $2.6\times
10^{-7}\,\mathrm{pc^{-3}}$ for the sample of 6 IPs, compared to our
best estimate of $3^{+2}_{-1}\times
10^{-7}\,\mathrm{pc^{-3}}$. Assuming $kT=30\,\mathrm{keV}$ for IPs,
but with either high ($N_H =4 \times 10^{22}\,\mathrm{cm^{-2}}$ and a
covering fraction of 1 for all systems) or no intrinsic absorption, we
find $\rho_0$ distributions with medians at $2.5\times
10^{-7}\,\mathrm{pc^{-3}}$ and $2.8\times 10^{-7}\,\mathrm{pc^{-3}}$,
respectively.

\begin{figure}
 \includegraphics[width=84mm]{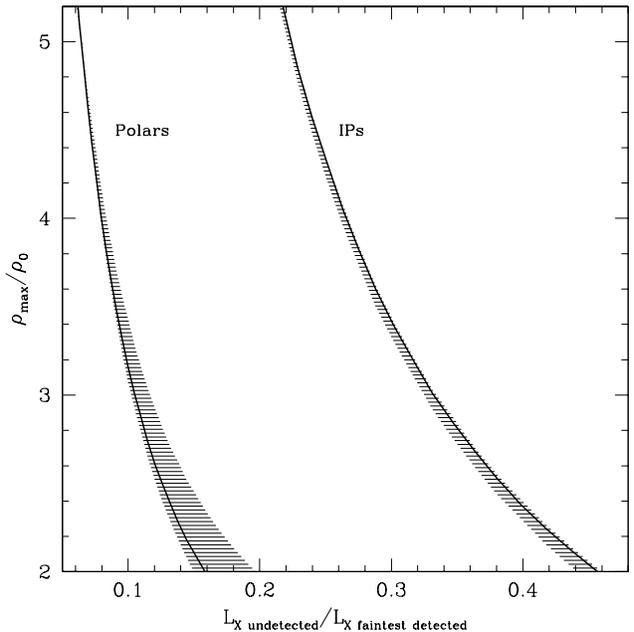} 
  \caption {$\rho_{max,IP}/\rho_{IP}$ and $\rho_{max,polar}/\rho_{polar}$ as a function of the ratios of the luminosity of a hidden population and the luminosity of the faintest detected IP/polar (CC Scl/EQ Cet). The bold curves are for our best model (a $30\,\mathrm{eV}$ blackbody spectrum for polars and partially absorbed $30\,\mathrm{keV}$ thermal bremsstrahlung spectra for IPs). The hatched areas show the sensitivity to the assumed spectra. For polars, this corresponds to the area between the curves obtained by assuming respectively $kT=10$ and $50\,\mathrm{eV}$ blackbody spectra in calculating $\rho_{max,polar}$ and X-ray luminosities for both the undetected population and EQ Cet (while $\rho_{polar}$ is fixed at $5\times 10^{-7}\,\mathrm{pc^{-3}}$). For IPs, the hatched area corresponds to $kT=10$ and $50\,\mathrm{keV}$ bremsstrahlung spectra. 
}
 \label{fig:limitratio}
\end{figure}

The upper limits on the space densities of hypothetical hidden
populations of IPs and polars are of course also affected by the
assumed X-ray spectrum. Since we find the upper limits as a function
of the X-ray luminosity of the undetected population (and $L_X$, for
the detected as well as ``hidden'' systems, depends on the assumed
X-ray spectrum), we present the effect of assuming different X-ray
spectra as the ratio of $\rho_{max}$ and the best-estimate measured
$\rho_0$, as a function of the ratio of $L_X$ of the hidden population
and the faintest detected system.  In other words, we fix
$\rho_{polar}=5\times 10^{-7}\,\mathrm{pc^{-3}}$ and
$\rho_{IP}=3\times 10^{-7}\,\mathrm{pc^{-3}}$, and then find
$\rho_{max,polar}$ and $\rho_{max,IP}$, as well as $L_X$ of the hidden
population of IPs and polars and of our faintest observed IP and polar
with different assumed spectra.  The result is shown in
Fig.~\ref{fig:limitratio}. In the case of polars, a population with a
space density as high as $5 \times$ the measured $\rho_{polar}$ (i.e.,
$\rho_{max}\simeq 4\times 10^{-6}\,\mathrm{pc^{-3}}$ must have $L_X
\la 0.06 \times$ the luminosity of EQ Cet, the faintest polar in the
RBS sample. An undetected IP population must have $L_X \la 0.2 \times$
the luminosity of the faintest detected system (CC Scl) in order for
its space density to be as high as $5 \times \rho_{IP}$ (i.e.,
$\rho_{max}\simeq 1.4\times 10^{-6}\,\mathrm{pc^{-3}}$). These ratios
are not very sensitive to the spectrum we assume (see the size of the
hatched areas in Fig.~\ref{fig:limitratio}).

\section{Discussion}
\label{sec:discussion}

Having derived space density estimates for polars and IPs, let us
revisit the three key questions that provided the main motivation for
our study (see Sections~\ref{sec:intro} and~\ref{sec:context}). We
note from the outset that our goal here is merely to take a brief look
at some of the implications of our results; a full analysis of these
issues is well beyond the scope of the present paper.

\subsection{The relationship between intermediate polars and polars}

As noted in Section~\ref{sec:context}, a natural explanation for the
lack of short-$P_{orb}$ IPs is that long-period IPs evolve into polars
below the period gap. If one assumes that long-period IPs are the sole
progenitors of short-period polars, and that all IPs synchronize once
they have crossed the period gap, then the ratio of the space
densities of long-$P_{orb}$ IPs and short-$P_{orb}$ polars
($\rho_{IP,lp}$ and $\rho_{polar,sp}$) should simply reflect their
relative evolutionary time-scales. For the purpose of our qualitative
discussion, let us throw caution to the wind and assume that the
numbers in Table~\ref{tab:res} are reasonable approximations to the
total space densities, even though our samples are flux-limited and
even though polars and IPs have quite different X-ray luminosity
functions.

We then find that the observed logarithm of this ratio is
$\log{(\rho_{polar,sp}/\rho_{IP,lp})} = 0.32 \pm 0.36$ (this becomes
$\log{(\rho_{polar,sp}/\rho_{IP,lp})} = 0.63 \pm 0.36$, if we assume a
0.5 high-state duty cycle for polars). The ratio itself is therefore
$\simeq 2$, but this is accurate to only about a factor of two. If the
evolution of long-period IPs is really driven by MB, while that of
short-period polars is driven solely by GR, the evolutionary
time-scale of the latter is expected to exceed that of the former by
at least a factor of $\gtappeq 5$ (e.g. \citealt{KBP11}). This is
larger than the ratio of the inferred space densities, but still
completely consistent with it, given the rather large statistical
errors. In fact, at 2-$\sigma$, the uncertainties are large enough to
encompass both ratios exceeding 10 and ratios less than unity. This
means that, with the currently available space density estimates for
polars and IPs, we cannot place strong constraints on the evolutionary
relationship between the two classes. Nevertheless, it is interesting
to note that the simplest possible model, in which short-period polars
derive from long-period IPs, is not ruled out by their observed space
densities.

\subsection{The intrinsic fraction of magnetic CVs}

We can also combine our measurement of the space density of mCVs
(polars as well as IPs) with that of non-magnetic CVs
(\citealt{nonmagphi,NEPrho}) in order to obtain an estimate of the
intrinsic fraction of mCVs amongst the Galactic CV population
($f_{mCV}$). This again involves throwing caution to the wind to some
extent, since the samples involved are flux-limited, and the
populations being compared have different X-ray luminosity
functions. Thus it is quite possible that our estimate of the mCV
fraction is still affected by selection biases.

Keeping this caveat in mind, we find that $\log(f_{mCV}) =
-0.80^{+0.27}_{-0.36}$, i.e $f_{mCV} \simeq 16$\%, to within a factor
of 2 (or $\log(f_{mCV}) = -0.63^{+0.24}_{-0.33}$, if half of all
polars are in the low state at a given time, and thus undetected in
the RBS).  This is consistent with the raw incidence of mCVs in the
known CV sample ($\simeq$20\%; see \citealt{rkcat}). However, more
importantly, it is also consistent, within our considerable
uncertainties, with the fraction of isolated WDs that are strongly
magnetic ($\sim 10$\%).  In fact, it seems likely that our
X-ray-selected CV sample is more complete for mCVs than it is for
non-magnetic CVs. If so, our estimate of $f_{mCV}$ should be
considered as an upper limit. Thus the incidence of magnetism is not
obviously enhanced amongst CV primaries compared to isolated WDs.

\subsection{Galactic X-ray Source Populations}
\label{sec:xraypop}
As noted in Section~\ref{sec:context}, mCVs -- and particularly IPs --
have been suggested to be the dominant X-ray source populations above
$L_X \simeq 10^{31} {\rm erg s^{-1}}$ in a variety of Galactic
environments. We can use our new space density estimates to check
whether mCVs can plausibly account for the number of sources seen in
surveys of these environments.

Let us take the Galactic Centre region as an example. The deep Chandra
survey of \cite{Muno09} covers an effective area of $\simeq 10^{-3}
{\rm deg^{2}}$ down to $L_X \simeq 10^{31}\,{\rm erg s^{-1}}$ and
includes $\simeq 9000$ sources. For our order of magnitude estimate
here, we will ignore subtleties like the flux/luminosity-dependent
survey area and simply ask if it is plausible that the majority of
these sources may be IPs.

Given that the stellar density distribution is highly peaked towards
the Galactic Centre, let us approximate the volume covered by the
survey as a sphere of radius $R \simeq 150$~pc. The space density of
X-ray sources in the Galactic Centre is then of order $\rho_{X,GC}
\sim 6\times 10^{-4}\,{\rm pc^{-3}}$, while the local space density of
IPs is $\rho_{IP} \sim 3\times 10^{-7}\,{\rm pc^{-3}}$. However, the
stellar space density in the Galactic centre is $\simeq 70\,{\rm
  pc^{-3}}$, while it is only $\simeq 0.044\,{\rm pc^{-3}}$ in the
solar neighborhood (e.g.\ \citealt{Hong09}).  We thus find that there
is roughly 1 X-ray source per 100,000 stars in the Galactic centre,
and roughly 1 IP per 200,000 stars in the solar neighborhood. At the
level of precision to which we are working here, these numbers are
identical. We thus conclude that IPs remain a viable explanation for
most of the X-ray sources seen in the Galactic Centre.

It should be obvious that the calculation above is not to be taken too
seriously. Its purpose is merely to illustrate how our measurement of
the space density of mCVs, in general, and IPs, in particular, relates
to recent X-ray surveys in a wide variety of Galactic environments. A
correct analysis would have to account in much more detail for the
properties of the various surveys. This is worth doing, but beyond the
scope of the present paper.

\subsection{Outlook}
The rough calculations above show that open questions concerning the
evolution and Galactic abundance of mCVs cannot yet be conclusively
answered.  This highlights a fundamental problem: given the limited
size of the existing flux-limited CV samples, and the low precision of
most available distance estimates, it is currently impossible to
measure space densities to an accuracy much better than a factor of
$\simeq 2$. To make matters worse, we often need ratios of space
densities for various sub-populations in order to test evolutionary
models, which necessarily suffer from even lower precision.

This situation should improve dramatically over the coming
years. Surveys with \emph{eROSITA} will reach flux limits around 2
orders of magnitude deeper than the RBS and will yield large
X-ray-selected mCV samples (e.g.\ \citealt{Schwope12}), while {\it
  Gaia} will provide accurate distance measurements for a large number
of CVs.

\section{Conclusions}
\label{sec:concl}
We have used a complete, purely X-ray flux-limited sample of 30 mCVs
from the RBS to place constraints on the local space density of these
systems. Our conclusions are listed below.
\begin{enumerate}
\item Assuming that the sample used here is representative of the
  intrinsic population (in the sense that the RBS detected at least
  one IP and one polar at the faintest ends of the luminosity
  functions of those populations), we obtain a mid-plane space density
  of $8^{+4}_{-2} \times 10^{-7}\,\mathrm{pc^{-3}}$ for mCVs. For the
  two distinct types of mCVs, we find $\rho_{polar}=5^{+3}_{-2} \times
  10^{-7}\,\mathrm{pc^{-3}}$ and $\rho_{IP}=3^{+2}_{-1} \times
  10^{-7}\,\mathrm{pc^{-3}}$. If we assume that polars are detectable
  in X-rays for only 50\% of the time, we obtain
  $\rho_{polar}=1^{+0.5}_{-0.3} \times 10^{-6}\,\mathrm{pc^{-3}}$ and
  $\rho_{mCV}=1.3^{+0.6}_{-0.6} \times 10^{-6}\,\mathrm{pc^{-3}}$.
\item We have calculated the maximum sizes of hypothetical faint
  populations of IPs and polars that are consistent with their
  non-detection in the RBS, as a function of the X-ray luminosity of
  the undetected populations.  If an undetected population of polars
  with a space density $5 \times$ as high as the space density we
  infer from detected systems exists, then those systems must have
  $L_X \la 0.06 \times$ the luminosity of the faintest polar in the
  RBS sample (EQ Cet). In the case of IPs, an undetected population
  with $L_X \la 0.2 \times$ the luminosity of the faintest detected
  system (CC Scl) can have a space density as high as $5 \times$ the
  value we measure from the detected IPs.
\item The ratio of the space density of short-period polars to
  long-period IPs is $2^{+3}_{-1}$ (or $4^{+6}_{-2}$, assuming a 50\%
  high-state duty cycle for polars). Within the large errors, this is
  consistent with the very simple hypothesis that (the majority of)
  long-period IPs evolve into short-period polars, and that this
  accounts for the whole population of short-period polars.
\item Our estimate of the intrinsic fraction of mCVs is consistent
  with the fraction of magnetic systems in the known CV sample
  ($\simeq 0.2$). However, with existing data, this fraction cannot be
  measured to high enough precision to rule out an incidence of mCVs
  as low as $\simeq$10\%. It is therefore not clear whether the
  fraction of strongly magnetic WDs is higher in CVs than in the
  single WD population.
\item When the local space density of IPs is scaled to the density of
  stars in the Galactic Centre, it is sufficiently high to account for
  the number of bright ($L_X \ga 10^{31}\,{\rm erg s^{-1}}$) X-ray
  sources detected in that region.
\end{enumerate}

\section*{Acknowledgements}
We thank Kars Verbeek for taking snapshot IR images of V4738 Sgr and RS Cae.

\bsp

\label{lastpage}


\begin{thebibliography}{}
\bibitem[\protect\citeauthoryear{Ak et al.}{2008}]{Ak08} Ak T., Bilir S., Ak S., Eker Z., 2008, NewA, 13, 133 
\bibitem[\protect\citeauthoryear{Am{\^o}res \& L{\'e}pine}{2005}]{AmoresLepine05} Am{\^o}res E.~B., L{\'e}pine J.~R.~D., 2005, AJ, 130, 659 
\bibitem[\protect\citeauthoryear{Araujo-Betancor et al.}{2005}]{Araujo-BetancorGansickeLong05} Araujo-Betancor S., G{\"a}nsicke B.T., Long K.S., Beuermann K., de Martino D., Sion E.M., Szkody P., 2005, ApJ, 622, 589 
\bibitem[\protect\citeauthoryear{Bailey}{1981}]{Bailey81} Bailey J., 1981, MNRAS, 197, 31 
\bibitem[\protect\citeauthoryear{Barrett, Singh \& Mitchell}{1999}]{BarrettSinghMitchell99} Barrett P., Singh K.~P., Mitchell S., in Hellier C., Mukai K., eds, ASP Conf. Ser. Vol. 157, Annapolis Workshop on Magnetic Cataclysmic Variables. Astron. Soc. Pac., San Francisco, p. 180
\bibitem[\protect\citeauthoryear{Beuermann}{2000}]{Beuermann00} Beuermann, K.\ 2000, NewAR, 44, 93 
\bibitem[\protect\citeauthoryear{Beuermann \& Burwitz}{1995}]{BeuermannBurwitz95}Beuermann K., Burwitz V., 1995, ASPC, 85, 99 
\bibitem[\protect\citeauthoryear{Beuermann \& Schwope}{1994}]{BeuermannSchwope94} Beuermann K., Schwope A.~D., 1994, in Shafter A. W., ed., ASP Conf. Ser. Vol. 56, Interacting Binary Stars. Astron. Soc. Pac., San Francisco p. 119
\bibitem[\protect\citeauthoryear{Beuermann \& Thomas}{1993}]{BeuermannThomas93} Beuermann K., Thomas H.-C., 1993, AdSpR, 13, 115 
\bibitem[\protect\citeauthoryear{Beuermann, Burwitz \& Reinsch}{2012}]{BeuermannBurwitzReinsch12} Beuermann K., Burwitz V., Reinsch K., 2012, A\&A, 543, A41
\bibitem[\protect\citeauthoryear{Beuermann, Thomas \& Pietsch}{1991}]{BeuermannThomasPietsch91} Beuermann K., Thomas H.-C., Pietsch W., 1991, A\&A, 246, L36 
\bibitem[\protect\citeauthoryear{Beuermann et al.}{1985}]{BeuermannSchwopeWeissieker85} Beuermann K., Schwope A., Weissieker H., Motch C., 1985, SSRv, 40, 135 
\bibitem[\protect\citeauthoryear{Beuermann et al.}{1999}]{BeuermannThomasReinsch99} Beuermann K., Thomas H.-C., Reinsch K., Schwope A.~D., Tr{\"u}mper J., Voges W., 1999, A\&A, 347, 47 
\bibitem[\protect\citeauthoryear{Beuermann et al.}{2003}]{Beuermann03} Beuermann K., Harrison T.~E., McArthur B.~E., Benedict G.~F., G{\"a}nsicke B.~T., 2003, A\&A, 412, 821 
\bibitem[\protect\citeauthoryear{Beuermann et al.}{2009}]{BeuermannDiesePaik09} Beuermann, K., Diese, J., Paik, S., Ploch, A., Zachmann, J., Schwope, A.~D., \& Hessman, F.~V.\ 2009, \aap, 507, 385
\bibitem[\protect\citeauthoryear{Bonnet-Bidaud et al.}{1996}]{Bonnet-Bidaud96} Bonnet-Bidaud, J.~M., Mouchet, M., Somova, T.~A., \& Somov, N.~N.\ 1996, \aap, 306, 199
\bibitem[\protect\citeauthoryear{Breedt et al.}{2012}]{Breedt12} Breedt E., G{\"a}nsicke B.~T., Girven J., Drake A.~J., Copperwheat C.~M., Parsons S.~G., Marsh T.~R., 2012, MNRAS, 423, 1437 
\bibitem[\protect\citeauthoryear{Brinkworth et al.}{2007}]{Brinkworth07} Brinkworth C.~S., et al., 2007, ApJ, 659, 1541 
\bibitem[\protect\citeauthoryear{Bruch \& Engel}{1994}]{BruchEngel94} Bruch A., Engel A., 1994, A\&AS, 104, 79 
\bibitem[\protect\citeauthoryear{Burwitz et al.}{1996}]{BurwitzReinschSchwope96} Burwitz V., Reinsch K., Schwope A.~D., Beuermann K., Thomas H.-C., Greiner J., 1996, A\&A, 305, 507 
\bibitem[\protect\citeauthoryear{Burwitz et al.}{1997}]{BurwitzReinschBeuermann97} Burwitz V., Reinsch K., Beuermann K., Thomas H.-C., 1997, A\&A, 327, 183 
\bibitem[\protect\citeauthoryear{Burwitz et al.}{1998}]{Burwitz98} Burwitz V., et al., 1998, A\&A, 331, 262 
\bibitem[\protect\citeauthoryear{Butters et al.}{2009}]{Butters09a} Butters O.~W., Katajainen S., Norton A.~J., Lehto H.~J., Piirola V., 2009, A\&A, 496, 891 
\bibitem[\protect\citeauthoryear{Cao, Wei \& Hu}{1999}]{CaoWeiHu99} Cao L., Wei J.-Y., Hu J.-Y., 1999, A\&AS, 135, 243 
\bibitem[\protect\citeauthoryear{Chanmugam \& Ray}{1984}]{ChanmugamRay84} Chanmugam G., Ray A., 1984, ApJ, 285, 252 
\bibitem[\protect\citeauthoryear{Chen et al.}{2001}]{ChenODonoghueStobie01} Chen A., O'Donoghue D., Stobie R.~S., Kilkenny D., Warner B., 2001, MNRAS, 325, 89 
\bibitem[\protect\citeauthoryear{Ciardi et al.}{1998}]{Ciardi98} Ciardi, D.~R., Howell, S.~B., Hauschildt, P.~H., \& Allard, F.\ 1998, \apj, 504, 450 
\bibitem[\protect\citeauthoryear{Clayton \& Osborne}{1994}]{ClaytonOsborne94} Clayton, K.~L., \& Osborne, J.~P.\ 1994, \mnras, 268, 229 
\bibitem[\protect\citeauthoryear{Cropper}{1990}]{Cropper90} Cropper M., 1990, SSRv, 54, 195 
\bibitem[\protect\citeauthoryear{Cumming}{2002}]{Cumming02} Cumming A., 2002, MNRAS, 333, 589
 \bibitem[\protect\citeauthoryear{de Martino et al.}{2004}]{deMartino04} de Martino D., Matt G., Belloni T., Haberl F., Mukai K., 2004, A\&A, 415, 1009 
\bibitem[\protect\citeauthoryear{Drimmel, Cabrera-Lavers \& L{\'o}pez-Corredoira}{Drimmel et~al.}{2003}]{Drimmel03} Drimmel R., Cabrera-Lavers A., L{\'o}pez-Corredoira M., 2003, \aap, 409, 205
\bibitem[\protect\citeauthoryear{Frank, King \& Raine}{1985}]{FKR85} Frank J., King A.~R., Raine D.~J., 1985, Accretion Power in Astrophysics, Cambridge Univ. Press, Cambridge
\bibitem[\protect\citeauthoryear{Frisch, Redfield \& Slavin}{2011}]{FrischRedfieldSlavin11} Frisch P.~C., Redfield S., Slavin J.~D., 2011, ARA\&A, 49, 237
\bibitem[\protect\citeauthoryear{G{\"a}nsicke}{1999}]{Gansicke99} G{\"a}nsicke B.~T., 1999, in Hellier C., Mukai K., eds, ASP Conf. Ser. 157, Proc. Annapolis Workshop on Magnetic Cataclysmic Variables. Astron. Soc. Pac., San Francisco, p. 261
\bibitem[\protect\citeauthoryear{G{\"a}nsicke et al.}{2009}]{Gansicke09} G{\"a}nsicke B.~T., et al., 2009, MNRAS, 397, 2170 
\bibitem[\protect\citeauthoryear{Gioia et al.}{2003}]{Gioia03} Gioia I.M., Henry J.P., Mullis C.R., B{\"o}hringer H., Briel U.G., Voges W., Huchra J.P., 2003, ApJS, 149, 29 
\bibitem[\protect\citeauthoryear{Glass \& Nagata}{2000}]{GlassNagata00} Glass I.~S., Nagata T., 2000, MNSSA, 59, 110 
\bibitem[\protect\citeauthoryear{Glenn et al.}{1994}]{Glenn94} Glenn J., Howell S.~B., Schmidt G.~D., Liebert J., Grauer A.~D., Wagner R.~M., 1994, ApJ, 424, 967 
\bibitem[\protect\citeauthoryear{Greiner, Schwarz \& Wenzel}{1998}]{GreinerSchwarzWenzel98} Greiner J., Schwarz R., Wenzel W., 1998, MNRAS, 296, 437 
\bibitem[\protect\citeauthoryear{Haberl \& Motch}{1995}]{HaberlMotch95} Haberl, F., \& Motch, C.\ 1995, \aap, 297, L37 
\bibitem[\protect\citeauthoryear{Heinke et al.}{2005}]{Heinke05} Heinke C.~O., Grindlay J.~E., Edmonds P.~D., Cohn H.~N., Lugger P.~M., Camilo F., Bogdanov S., Freire P.~C., 2005, ApJ, 625, 796 
\bibitem[\protect\citeauthoryear{Hellier}{1993}]{Hellier93}Hellier C., 1993, MNRAS, 264, 132 
\bibitem[\protect\citeauthoryear{Henry et al.}{2006}]{Henry06} Henry J.P., Mullis C.R., Voges W., B{\"o}hringer H., Briel U.G., Gioia I.M., Huchra J.P., 2006, ApJS, 162, 304 
\bibitem[\protect\citeauthoryear{Hessman, G{\"a}nsicke, \& Mattei}{2000}]{Hessman00} Hessman F.~V., G{\"a}nsicke B.~T., Mattei J.~A., 2000, A\&A, 361, 952 
\bibitem[\protect\citeauthoryear{Hoard et al.}{2005}]{HoardLinnellSzkody05} Hoard D.~W., Linnell A.~P., Szkody P., Sion E.~M., 2005, AJ, 130, 214 
\bibitem[\protect\citeauthoryear{Hong et al.}{2009}]{Hong09} Hong J.~S., van den Berg M., Grindlay J.~E., Laycock S., 2009, ApJ, 706, 223 
\bibitem[\protect\citeauthoryear{Hong et al.}{2012}]{Hong12} Hong J., van den Berg M., Grindlay J.~E., Servillat M., Zhao P., 2012, ApJ, 746, 165 
\bibitem[\protect\citeauthoryear{Howell et al.}{1995}]{Howell95} Howell S.~B., Sirk M.~M., Malina R.~F., Mittaz J.~P.~D., Mason K.~O., 1995, ApJ, 439, 991
\bibitem[\protect\citeauthoryear{Kalberla et al.}{2005}]{Kalberla05} Kalberla P.~M.~W., Burton W.~B., Hartmann D., Arnal E.~M., Bajaja E., Morras R., P{\"o}ppel W.~G.~L., 2005, A\&A, 440, 775
\bibitem[\protect\citeauthoryear{Katajainen et al.}{2000}]{Katajainen00} Katajainen S., Lehto H.~J., Piirola V., Karttunen H., Piironen J., 2000, A\&A, 357, 677
\bibitem[\protect\citeauthoryear{Kawka et al.}{2007}]{Kawka07} Kawka A., Vennes S., Schmidt G.~D., Wickramasinghe D.~T., Koch R., 2007, ApJ, 654, 499 
\bibitem[\protect\citeauthoryear{King, Frank \& Ritter}{1985}]{KingFrankRitter85} King A.~R., Frank J., Ritter H., 1985, MNRAS, 213, 181 
\bibitem[\protect\citeauthoryear{Knigge}{2006}]{Knigge06} Knigge C., 2006, MNRAS, 373, 484 
\bibitem[\protect\citeauthoryear{Knigge, Baraffe \& Patterson}{Knigge et al.}{2011}]{KBP11} Knigge C., Baraffe I., Patterson J., 2011, ApJS, 194, 28
\bibitem[\protect\citeauthoryear{Krivonos et al.}{2007}]{Krivonos07} Krivonos R., Revnivtsev M., Churazov E., Sazonov S., Grebenev S., Sunyaev R., 2007, A\&A, 463, 957 
\bibitem[\protect\citeauthoryear{Krzeminski \& Serkowski}{1977}]{KrzeminskiSerkowski77} Krzeminski W., Serkowski K., 1977, ApJ, 216, L45 
\bibitem[\protect\citeauthoryear{K{\"u}lebi et al.}{2009}]{Kulebi09} K{\"u}lebi B., Jordan S., Euchner F., G{\"a}nsicke B.~T., Hirsch H., 2009, A\&A, 506, 1341 
\bibitem[\protect\citeauthoryear{La Dous}{1991}]{LaDous91} La Dous C., 1991, A\&A, 252, 100 
\bibitem[\protect\citeauthoryear{Lamb \& Melia}{1987}]{LambMelia87} Lamb D.~Q., Melia F., 1987, Ap\&SS, 131, 511 
\bibitem[\protect\citeauthoryear{Li \& Wickramasinghe}{1998}]{LiWickramasinghe98} Li J., Wickramasinghe D.~T., 1998, MNRAS, 300, 718 
\bibitem[\protect\citeauthoryear{Liebert et al.}{1982}]{Liebert82} Liebert J., Tapia S., Bond H.~E., Grauer A.~D., 1982, ApJ, 254, 232 
\bibitem[\protect\citeauthoryear{Lutz \& Kelker}{1973}]{LKbias} Lutz T.~E., Kelker D.~H., 1973, PASP, 85, 573 
\bibitem[\protect\citeauthoryear{Malmquist}{1924}]{Mbias}Malmquist, K.G. 1924, Medd. Lund Astron. Obs., 2(32), 64 
\bibitem[\protect\citeauthoryear{Mateo, Szkody, \& Garnavich}{1991}]{MateoSzkodyGarnavich91} Mateo M., Szkody P., Garnavich P., 1991, ApJ, 370, 370 
\bibitem[\protect\citeauthoryear{McArthur et al.}{2001}]{McArthur01} McArthur B.~E., et al., 2001, ApJ, 560, 907 
 \bibitem[\protect\citeauthoryear{Mittaz et al.}{1992}]{Mittaz92} Mittaz J.~P.~D., Rosen S.~R., Mason K.~O., Howell S.~B., 1992, MNRAS, 258, 277 
\bibitem[\protect\citeauthoryear{Monet et al.}{2003}]{Monet03} Monet D.~G., et al., 2003, AJ, 125, 984 
\bibitem[\protect\citeauthoryear{Morris et al.}{1987}]{Morris87} Morris S.~L., Schmidt G.~D., Liebert J., Stocke J., Gioia I.~M., Maccacaro T., 1987, ApJ, 314, 641 
\bibitem[\protect\citeauthoryear{Mouchet et al.}{1991}]{Mouchet91} Mouchet M., Bonnet-Bidaud J.~M., Buckley D.~A.~H., Tuohy I.~R., 1991, A\&A, 250, 99 
\bibitem[\protect\citeauthoryear{Mukai et al.}{2003}]{Mukai03} Mukai K., Kinkhabwala A., Peterson J.~R., Kahn S.~M., Paerels F., 2003, ApJ, 586, L77 
\bibitem[\protect\citeauthoryear{Muno et al.}{2004}]{Muno04} Muno M.~P., et al., 2004, ApJ, 613, 1179 
\bibitem[\protect\citeauthoryear{Muno et al.}{2006}]{Muno06} Muno M.~P., Bauer F.~E., Bandyopadhyay R.~M., Wang Q.~D., 2006, ApJS, 165, 173 
\bibitem[\protect\citeauthoryear{Muno et al.}{2009}]{Muno09}Muno M.~P., et al., 2009, ApJS, 181, 110 
\bibitem[\protect\citeauthoryear{Nagayama et al.}{2003}]{Nagayama03} Nagayama T., et al., 2003, SPIE, 4841, 459 
\bibitem[\protect\citeauthoryear{Norton \& Watson}{1989}]{NortonWatson89} Norton, A.~J., \& Watson, M.~G.\ 1989, \mnras, 237, 853 
\bibitem[\protect\citeauthoryear{Norton, Wynn \& Somerscales}{2004}]{NortonWynnSomerscales04} Norton A.~J., Wynn G.~A., Somerscales R.~V., 2004, ApJ, 614, 349 
\bibitem[\protect\citeauthoryear{Norton et al.}{2000}]{Norton00} Norton A.~J., Beardmore A.~P., Retter A., Buckley D.~A.~H., 2000, MNRAS, 312, 362 
\bibitem[\protect\citeauthoryear{Osborne et al.}{1994}]{OsborneBeardmoreWheatley94} Osborne J.~P., Beardmore A.~P., Wheatley P.~J., Hakala P., Watson M.~G., Mason K.~O., Hassall B.~J.~M., King A.~R., 1994, MNRAS, 270, 650 
\bibitem[\protect\citeauthoryear{Pandel \& C{\'o}rdova}{2005}]{PandelCordova05} Pandel D., C{\'o}rdova F.~A., 2005, ApJ, 620, 416 
\bibitem[\protect\citeauthoryear{Patterson}{1984}]{Patterson84} Patterson J., 1984, ApJS, 54, 443 
\bibitem[\protect\citeauthoryear{Patterson}{1994}]{Patterson94} Patterson J., 1994, PASP, 106, 209 
\bibitem[\protect\citeauthoryear{Patterson}{2011}]{Patterson11} Patterson J., 2011, MNRAS, 411,2695
\bibitem[\protect\citeauthoryear{Patterson \& Moulden}{1993}]{PattersonMoulden93} Patterson J., Moulden M., 1993, PASP, 105, 779 
\bibitem[\protect\citeauthoryear{Patterson \& Price}{1981}]{PattersonPrice81} Patterson J., Price C.~M., 1981, ApJ, 243, L83 
\bibitem[\protect\citeauthoryear{Patterson et al.}{2005}]{Patterson05} Patterson J., et al., 2005, PASP, 117, 1204 
\bibitem[\protect\citeauthoryear{Peters}{2008}]{Petersthesis} Peters C.~S., 2008, PhD thesis, Dartmouth College
\bibitem[\protect\citeauthoryear{Pietsch et al.}{1987}]{Pietsch87} Pietsch W., Voges W., Kendziorra E., Pakull M., 1987, Ap\&SS, 130, 281 
\bibitem[\protect\citeauthoryear{Predehl \& Schmitt}{1995}]{PredehlSchmitt95} Predehl P., Schmitt J.H.M.M., 1995, A\&A, 293, 889 
\bibitem[\protect\citeauthoryear{Pretorius \& Knigge}{2012}]{nonmagphi} Pretorius M.~L., Knigge C., 2012, MNRAS, 419, 1442
\bibitem[\protect\citeauthoryear{Pretorius, Knigge \& Kolb}{Pretorius et al.}{2007a}]{PretoriusKniggeKolb07} Pretorius M.~L., Knigge C., Kolb U., 2007a, MNRAS, 374, 1495 
\bibitem[\protect\citeauthoryear{Pretorius et al.}{2007b}]{NEPrho}Pretorius M.~L., Knigge C., O'Donoghue D., Henry J.~P., Gioia I.~M., Mullis C.~R., 2007b, MNRAS, 382, 1279 
\bibitem[\protect\citeauthoryear{Ramsay \& Cropper}{2004}]{RamsayCropper04} Ramsay G., Cropper M., 2004, MNRAS, 347, 497 
\bibitem[\protect\citeauthoryear{Ramsay \& Cropper}{2007}]{RamsayCropper07} Ramsay G., Cropper M., 2007, MNRAS, 379, 1209 
\bibitem[\protect\citeauthoryear{Ramsay, Cropper \& Mason}{1996}]{RamsayCropperMason96} Ramsay G., Cropper M., Mason K.~O., 1996, MNRAS, 278, 285 
\bibitem[\protect\citeauthoryear{Ramsay et al.}{1994}]{Ramsay94} Ramsay G., Mason K.~O., Cropper M., Watson M.~G., Clayton K.~L., 1994, MNRAS, 270, 692
\bibitem[\protect\citeauthoryear{Ramsay et al.}{1999}]{Ramsay99} Ramsay G., Buckley D.~A.~H., Cropper M., Harrop-Allin M.~K., 1999, MNRAS, 303, 96 
\bibitem[\protect\citeauthoryear{Ramsay et al.}{2004a}]{RamsayCropperMason04} Ramsay, G., Cropper, M., Mason, K.~O., C{\'o}rdova, F.~A., \& Priedhorsky, W.\ 2004a, \mnras, 347, 95 
\bibitem[\protect\citeauthoryear{Ramsay et al.}{2004b}]{Ramsay04} Ramsay G., Cropper M., Wu K., Mason K.~O., C{\'o}rdova F.~A., Priedhorsky W., 2004b, MNRAS, 350, 1373 
\bibitem[\protect\citeauthoryear{Ramsay et al.}{2008}]{Ramsay08} Ramsay G., Wheatley P.~J., Norton A.~J., Hakala P., Baskill D., 2008, MNRAS, 387, 1157 
\bibitem[\protect\citeauthoryear{Ramsay et al.}{2009}]{Ramsay09} Ramsay G., Rosen S., Hakala P., Barclay T., 2009, MNRAS, 395, 416 
\bibitem[\protect\citeauthoryear{Rana et al.}{2004}]{Rana04} Rana, V.~R., Singh, K.~P., Schlegel, E.~M., \& Barrett, P.\ 2004, \aj, 127, 489 
\bibitem[\protect\citeauthoryear{Reimer et al.}{2008}]{Reimer08} Reimer T.~W., Welsh W.~F., Mukai K., Ringwald F.~A., 2008, ApJ, 678, 376 
\bibitem[\protect\citeauthoryear{Reinsch et al.}{1994}]{ReinschBurwitzBeuermann94} Reinsch K., Burwitz V., Beuermann K., Schwope A.~D., Thomas H.-C., 1994, A\&A, 291, L27 
\bibitem[\protect\citeauthoryear{Remillard et al.}{1994}]{Remillard94} Remillard, R.~A., Schachter, J.~F., Silber, A.~D., Slane, P.\ 1994, \apj, 426, 288 
\bibitem[\protect\citeauthoryear{Revnivtsev et al.}{2006}]{Revnivtsev06} Revnivtsev M., Sazonov S., Gilfanov M., Churazov E., Sunyaev R., 2006, A\&A, 452, 169 
\bibitem[\protect\citeauthoryear{Revnivtsev et al.}{2008}]{Revnivtsev08} Revnivtsev M., Sazonov S., Krivonos R., Ritter H., Sunyaev R., 2008, A\&A, 489, 1121
\bibitem[\protect\citeauthoryear{Richman}{1996}]{Richman96} Richman H.~R., 1996, ApJ, 462, 404 
\bibitem[\protect\citeauthoryear{Ritter \& Kolb}{2003}]{rkcat}Ritter H., Kolb U., 2003, A\&A, 404, 301 (update RKcat7.18, 2012)
\bibitem[\protect\citeauthoryear{Rodrigues et al.}{2006}]{Rodrigues06} Rodrigues, C.~V., Jablonski, F.~J., D'Amico, F., Cieslinski, D., Steiner, J.~E., Diaz, M.~P., Hickel, G.~R.\ 2006, MNRAS, 369, 1972 
\bibitem[\protect\citeauthoryear{Romero-Colmenero et al.}{2003}]{Romero-Colmenero03} Romero-Colmenero E., Potter S.~B., Buckley D.~A.~H., Barrett P.~E., Vrielmann S., 2003, MNRAS, 339, 685 
\bibitem[\protect\citeauthoryear{Sazonov et al.}{2006}]{Sazonov06} Sazonov S., Revnivtsev M., Gilfanov M., Churazov E., Sunyaev R., 2006, A\&A, 450, 117
\bibitem[\protect\citeauthoryear{Schmidt et al.}{1996}]{SchmidtSzkodySmith96} Schmidt G.~D., Szkody P., Smith P.~S., Silber A., Tovmassian G., Hoard D.~W., G{\"a}nsicke B.~T., de Martino D., 1996, ApJ, 473, 483 
\bibitem[\protect\citeauthoryear{Schmidt et al.}{2001}]{SchmidtFerrarioWickramasinghe01} Schmidt G.~D., Ferrario L., Wickramasinghe D.~T., Smith P.~S., 2001, ApJ, 553, 823 
\bibitem[\protect\citeauthoryear{Schwarz et al.}{1998}]{Schwarz98} Schwarz R., et al., 1998, A\&A, 338, 465 
\bibitem[\protect\citeauthoryear{Schwarz et al.}{2009}]{Schwarz09} Schwarz R., Schwope A.~D., Vogel J., Dhillon V.~S., Marsh T.~R., Copperwheat C., Littlefair S.~P., Kanbach G., 2009, A\&A, 496, 833
\bibitem[\protect\citeauthoryear{Schwope}{2012}]{Schwope12} Schwope A., 2012, in Giovanelli F., Sabau-Graziati L., eds, The golden age of cataclysmic variables and related objects, Mem. S. A. It. Vol. 83, p. 844
\bibitem[\protect\citeauthoryear{Schwope, Schreiber \& Szkody}{2006}]{SchwopeSchreiberSzkody06} Schwope A.~D., Schreiber M.~R., Szkody P., 2006, A\&A, 452, 955 
\bibitem[\protect\citeauthoryear{Schwope, Schwarz, \& Greiner}{1999}]{SchwopeSchwarzGreiner99} Schwope A.~D., Schwarz R., Greiner J., 1999, A\&A, 348, 861
\bibitem[\protect\citeauthoryear{Schwope, Thomas, \& Beuermann}{1993}]{SchwopeThomasBeuermann93}Schwope A.~D., Thomas H.~C., Beuermann K., 1993, A\&A, 271, L25
\bibitem[\protect\citeauthoryear{Schwope et al.}{1997}]{SchwopeBuckleyODonoghue97} Schwope A.~D., Buckley D.~A.~H., O'Donoghue D., Hasinger G., Truemper J., Voges W., 1997, A\&A, 326, 195 
\bibitem[\protect\citeauthoryear{Schwope et al.}{2000}]{Schwope00} Schwope A., et al., 2000, AN, 321, 1 
\bibitem[\protect\citeauthoryear{Schwope et al.}{2002}]{SchwopeBrunnerBuckley02}Schwope A.D., Brunner H., Buckley D., Greiner J., Heyden K.v.d., Neizvestny S., Potter S., Schwarz R., 2002, A\&A, 396, 895 
\bibitem[\protect\citeauthoryear{Schwope et al.}{2007}]{Schwope07} Schwope A.~D., Staude A., Koester D., Vogel J., 2007, A\&A, 469, 1027 
\bibitem[\protect\citeauthoryear{Singh et al.}{1995}]{Singh95} Singh K.~P., et al., 1995, ApJ, 453, L95 
\bibitem[\protect\citeauthoryear{Skrutskie et al.}{2006}]{2mass} Skrutskie M.F., et al., 2006, AJ, 131, 1163 
\bibitem[\protect\citeauthoryear{Sproats, Howell, \& Mason}{1996}]{SproatsHowellMason96} Sproats L.~N., Howell S.~B., Mason K.~O., 1996, MNRAS, 282, 1211 
\bibitem[\protect\citeauthoryear{Sterken et al.}{1983}]{Sterken93} Sterken C., Vogt N., Freeth R., Kennedy H.~D., Page A.~A., Marino B.~F., Walker W.~S.~G., 1983, A\&A, 118, 325 
\bibitem[\protect\citeauthoryear{Stobie, Ishida \& Peacock}{Stobie et al.}{1989}]{StobieIshidaPeacock89} Stobie R.~S., Ishida K., Peacock J.~A., 1989, MNRAS, 238, 709 
\bibitem[\protect\citeauthoryear{Tappert, Augusteijn \& Maza}{Tappert et al.}{2004}]{TappertAugusteijnMaza04} Tappert C., Augusteijn T., Maza J., 2004, MNRAS, 354, 321
\bibitem[\protect\citeauthoryear{Thomas \& Beuermann}{1998}]{ThomasBeuermann98} Thomas H.-C., Beuermann K., 1998, in Breitschwerdt D., Freyberg M.J., Truemper J., eds, Lecture Notes in Physics Vol. 506, The Local Bubble and Beyond. Springer-Verlag, Berlin, p. 247
\bibitem[\protect\citeauthoryear{Thomas et al.}{1996}]{ThomasBeuermannSchwope96} Thomas H.-C., Beuermann K., Schwope A.~D., Burwitz V., 1996, A\&A, 313, 833 
\bibitem[\protect\citeauthoryear{Thomas et al.}{1998}]{Thomas98} Thomas H.-C., Beuermann K., Reinsch K., Schwope A.~D., Truemper J., Voges W., 1998, A\&A, 335, 467 
\bibitem[\protect\citeauthoryear{Thorstensen}{1986}]{Thorstensen86}Thorstensen, J.~R.\ 1986, AJ, 91, 940
\bibitem[\protect\citeauthoryear{Thorstensen}{2003}]{Thorstensen03} Thorstensen J.~R., 2003, AJ, 126, 3017 
\bibitem[\protect\citeauthoryear{Thorstensen, L{\'e}pine \& Shara}{2008}]{ThorstensenLepineShara08} Thorstensen J.~R., L{\'e}pine S., Shara M., 2008, AJ, 136, 2107 
\bibitem[\protect\citeauthoryear{Thorstensen et al.}{2009}]{Thorstensen09} Thorstensen J.~R., Schwarz R., Schwope A.~D., Staude A., Vogel J., Krumpe M., Kohnert J., Nebot G{\'o}mez-Mor{\'a}n A., 2009, PASP, 121, 465 
\bibitem[\protect\citeauthoryear{Tinney, Reid \& Mould}{Tinney et al.}{1993}]{TinneyReidMould93} Tinney C.~G., Reid I.~N., Mould J.~R., 1993, ApJ, 414, 254
\bibitem[\protect\citeauthoryear{Tout et al.}{2008}]{Tout08} Tout C.~A., Wickramasinghe D.~T., Liebert J., Ferrario L., Pringle J.~E., 2008, MNRAS, 387, 897
\bibitem[\protect\citeauthoryear{Townsley \& G{\"a}nsicke}{2009}]{TownsleyGansicke09} Townsley D.~M., G{\"a}nsicke B.~T., 2009, ApJ, 693, 1007 
\bibitem[\protect\citeauthoryear{Verbunt}{1987}]{Verbunt87} Verbunt F., 1987, A\&AS, 71, 339 
\bibitem[\protect\citeauthoryear{Vergely et al.}{1998}]{Vergely98} Vergely J.-L., Ferrero R.~F., Egret D., Koeppen J., 1998, A\&A, 340, 543 
\bibitem[\protect\citeauthoryear{Visvanathan, Bessell, \& Wickramasinghe}{1984}]{Visvanathan84} Visvanathan N., Bessell M.~S., Wickramasinghe D.~T., 1984, IAUC, 3923, 2 
\bibitem[\protect\citeauthoryear{Vogel et al.}{2008}]{Vogel08} Vogel J., Byckling K., Schwope A., Osborne J.~P., Schwarz R., Watson M.~G., 2008, A\&A, 485, 787 
\bibitem[\protect\citeauthoryear{Voges et al.}{1999}]{rosatbsc} Voges W., et al., 1999, A\&A, 349, 389 
\bibitem[\protect\citeauthoryear{Voges et al.}{2000}]{rosatfsc} Voges W., et al., 2000, IAUC, 7432, 3 
\bibitem[\protect\citeauthoryear{Warner}{1995}]{bible} Warner B., 1995, Cataclysmic Variable Stars. Cambridge Univ.\ Press, Cambridge
\bibitem[\protect\citeauthoryear{Wickramasinghe, Wu \& Ferrario}{1991}]{WickramasingheWuFerrario91} Wickramasinghe D.~T., Wu K., Ferrario L., 1991, MNRAS, 249, 460 
\bibitem[\protect\citeauthoryear{Williams et al.}{1979}]{Williams79} Williams, G., Johns, M., Price, C., Hiltner, A., Boley, F., Maker, S., \& Mook, D.\ 1979, \nat, 281, 48 
\bibitem[\protect\citeauthoryear{Woudt et al.}{2012}]{Woudt12} Woudt P.~A., et al., 2012, MNRAS, in press (arXiv:1208.5936)
\bibitem[\protect\citeauthoryear{Wright et al.}{2010}]{wise} Wright E.~L., et al., 2010, AJ, 140, 1868
\bibitem[\protect\citeauthoryear{Wu \& Kiss}{2008}]{WuKiss08} Wu K., Kiss L.~L., 2008, A\&A, 481, 433 

\end{thebibliography}
\end{document}